\documentclass[12pt]{iopart}

\usepackage{graphicx}
\usepackage{dcolumn}
\usepackage{bm}

\usepackage[utf8]{inputenc}
\usepackage[T1]{fontenc}
\usepackage{mathptmx}

\usepackage{gensymb}
\usepackage[english]{babel}
\usepackage{booktabs}
\usepackage{subcaption}
\usepackage{xcolor,colortbl}
\definecolor{Gray}{gray}{0.85}
\definecolor{LightCyan}{rgb}{0.88,1,1}
\definecolor{Green}{rgb}{0.5,1,0.5}

\begin{document}


\title[Self Sustained Thermally Induced Gas-Damped Oscillations of Bimetal Cantilevers]{Self Sustained Thermally Induced Gas-Damped Oscillations of Bimetal Cantilevers with Application to the Design of a New Pyroelectric Micro Energy Harvester}


\author{Tarek Gebrael$^\star$, Ali Kanj$^\star$, Daniel Farhat$^\dagger$, Mutasem Shehadeh$^\dagger$, and Issam Lakkis$^{\dagger\mathsection}$}
\address{ 
$\star$ Department of Mechanical Science and Engineering, University of Illinois at Urbana-Champaign, Urbana, Illinois, USA
}%
\address{%
$\dagger$ Department of Mechanical Engineering, American University of Beirut, Beirut, Lebanon
}%
\ead{$\mathsection$ ilakkis@alum.mit.edu}

\vspace{10pt}
\begin{indented}
\item[]September 2019
\end{indented}

\begin{abstract}
Low efficiency is the main drawback of many MEMS thermal energy harvesters. Recently, energy harvesting micro-devices that operate using the pyroelectric effect gained attention due to their potential superior performance. Operation of these devices is based on the cyclic motion of a pyroelectric capacitor that operates between a high temperature and a low temperature reservoirs. In this paper, we investigate the dynamics of oscillations of a pyroelectric capacitor self sustained by thermally actuated bimetal micro-cantilevers, a topic which is so far underinvestigated. In addition to highlighting key thermodynamic aspects of the operation, we explore conditions for self-sustained oscillations and discuss the viability of operation at the mechanical resonance frequency. The analysis is presented for a new design inspired by the device proposed in Refs.\cite{2011,2012}, where in contrast, our proposed design boasts the following features: The pyroelectric capacitor remains parallel to the heat reservoirs, by virtue of its symmetric support by two bimetallic cantilever beams; In addition, the cyclic operation of the device does not require physical contact, thus lowering the risk of mechanical failure; To adjust the damping force imparted by the surrounding gas, the thermal reservoirs are equipped with trenches. To study the dynamic operation of the device, we developed a physically based reduced order, yet accurate, model that accounts for the heat transfer between and within the different components, and for the various forces including the gas damping force. The model is embedded within an optimization algorithm to produce optimal designs over the range $26-38 \degree C$ of temperature difference between the two reservoirs. The corresponding range of harvested power density is 0.4-0.65 $mW/cm^2$. 
\end{abstract}

\vspace{2pc}

%
%
\submitto{\JPD}
%
\maketitle
%
%

\section{Introduction}\label{sectionIntroduction}
It has been reported that a substantial amount of the consumed energy in the USA is wasted as heat \cite{bowen2014pyroelectric,pandya2018pyroelectric}.
Therefore, considerable research effort has been directed towards harvesting energy from these waste sources. This in turn would offer a long lasting green
power and maintenance free devices. Most of the research in this domain has been focused on utilizing thermoelectric materials which rely on the spatial gradient of temperature ($dT/dx$).  This temperature gradient can be easily realized without the need for moving parts, which is a desirable design feature. 
Thermoelectric devices are, however, not suited for applications where the temperature is below 100 $\degree C$ \cite{pandya2018pyroelectric}.  Moreover, extensive research
on thermoelectric generators showed that their efficiency could not exceed 1-5\%\cite{2011}. In contrast, pyroelectric devices depend on the time variation
of temperature ($dT/dt$) for power conversion, which is realized by cyclic motion of the pyroelectric capacitor operating between a high and a low temperature
reservoirs.

It is worth noting that the efficiency of pyroelectric devices is greatly affected by their operation frequency. That is the closer the frequency to the mechanical
resonance frequency, the higher the device efficiency. The operation frequency in bulk devices is expected to be in the order of 1-5 Hz, which is orders of magnitude
below the resonance frequency. This is due to limitations in the diffusion-based heat transfer processes and/or others imposed by the actuation mechanism.
In order to increase the operation frequency, heat exchange between the pyroelectric capacitor and the thermal reservoirs should be sufficiently fast. This can be accomplished by reducing the gap separating the capacitor and the heat source (sink) to the sub-micron length scale \cite{fang2010harvesting}, causing the much
faster radiative heat transfer to be the dominant mode of heat transfer. It has also been proposed to use thin films as a simple mean to improve the heat transfer rate \cite{yang2012pyroelectric}.  Equally important for fast operation of the device is the speed of the actuation mechanism. Self sustained cyclic motion
of the pyroelectric plate using thermally actuated bimetal cantilevers is proposed in Refs.\cite{2012, 2011}. In this case, heat transfer by conduction along the bimetal cantilevers is a speed limiting mechanism. Alternatively, actuation through piezoelectric nano-pillars is proposed in Ref. \cite{yang2012pyroelectric}. Although fast, this actuation method requires electric input work for actuation which, for acceptable values of the overall efficiency (or net generated work), yields small displacements of the pyroelectric capacitor. This, in turn, keeps the distance between the plate and the heat source/sink relatively large, which limits the heat exchange rate.  This may explain why the operation frequency does not exceed 10 Hz which is well below the natural frequency of the device \cite{yang2012pyroelectric}.




Self sustained oscillations (SSO) of piezo-electrically actuated micro-cantilevers have been widely used in many MEMS devices such as photo-detectors \cite{lulec2016mems}, atomic force microscopy \cite{lulec2016mems} and sensors \cite{lavrik2004cantilever,arlett2011comparative,muriuki2003design}.
In these systems, the oscillator is usually made from a single metallic or semiconducting material and their vibratory behaviors in different environments (air, liquid and even vacuum)  are well understood \cite{puchades2011thermally}. However, the literature is lacking when it comes to the analysis of gas-damped SSO of bimetal cantilevers actuated by cyclic temperature variation, such as those encountered in pyroelectric microdevices. It is therefore of paramount importance for reliable design of such systems to investigate their dynamic behavior. It is well documented that at nano-microscale, the performance of microdevices with vibrating components is strongly affected by the viscous damping due to their interaction with the
surrounding fluid. In parallel plate MEMS capacitors, for example, the distance between the capacitor plates is usually minimized for improved efficiency \cite{nayfeh2003new}. This potentially may result in non-continuum fluid-solid interaction due to the formation of squeezed film damping. 

In this paper, we propose a new pyroelectric micro energy harvester. We use the proposed device to investigate the dynamics of thermally induced self sustained oscillations of bimetal micro-cantilevers using a physically-based reduced order, yet accurate, model that we developed for this purpose. The model accounts for the heat transfer between and within the different components, and for the various forces including the damping force induced by the squeeze gas film. We then employ this model within an optimization loop to arrive at nearly optimal dimensions of the proposed design over a range of the temperature difference between the high and low temperature reservoirs. The proposed design is inspired by the device proposed in Refs.\cite{2011,2012}, which consists of a proof mass supported by a bimaterial canitlever that serves at the same time as the pyroelectric capacitor. The proof mass which oscillates between the high and low temperature reservoirs serves as a thermal capacitor that
stores heat when it contacts the high temperature reservoir (HTR) and loses heat when it contacts the low temperature reservoir (LTR). The oscillation, which generates an electrical current in the pyroelectric capacitor, is actuated by the downward and upward bending of the bimorph cantilever as a result of heat transfer from the proof mass to the bimaterial cantilever and vice versa. Our proposed device was designed, in part, to overcome the following limitations in the device proposed in Refs.\cite{2011,2012}.
The asymmetrical shape of the device requires tilting of the proof mass so that contact with the reservoir walls occurs over a small area. This design limits the rate of heat transfer between the
proof mass and the thermal reservoirs due to the resulting high contact thermal resistance. In addition, direct contact over a large number of operation cycles may lead to failure\cite{walraven2003failure,van2003mems,patton2002failure}.
Moreover, heat gained by the proof mass upon contact with the HTR flows in the pyroelectric capacitor along its length. The resulting pyroelectric effect, which depends on the spatial distribution of the temperature
variation with time decreases along the bimetal cantilever length away from the capacitor.

This paper is organized as follows. The proposed design is presented in Section \ref{sec:theProposedDesign}. In Section \ref{sec:Thermodynamics_Model}, we highlight key thermodynamic aspects of the operation. The reduced order model we developed to
design the device and investigate its dynamic operation is presented in
Section \ref{sectionReducedOrderModel}.
Dynamic operation of the device is investigated in Section
\ref{sec:dynamicOperation}. In Section \ref{sec:designConsiderationsAndOptimization},
we outline key design considerations, discuss conditions for sustained oscillations,
and present a simple design procedure for device operation at the mechanical resonance frequency. Assessment of the validity and accuracy of the model is conducted in Section \ref{modelValidation}.
Optimized designs and trends characterizing key performance indicators
including the energy output are presented in Section \ref{optimalDesigns}.


%

\section{The Proposed Device}\label{sec:theProposedDesign}
\begin{figure}[!ht]
\centering
\includegraphics[width=4in]{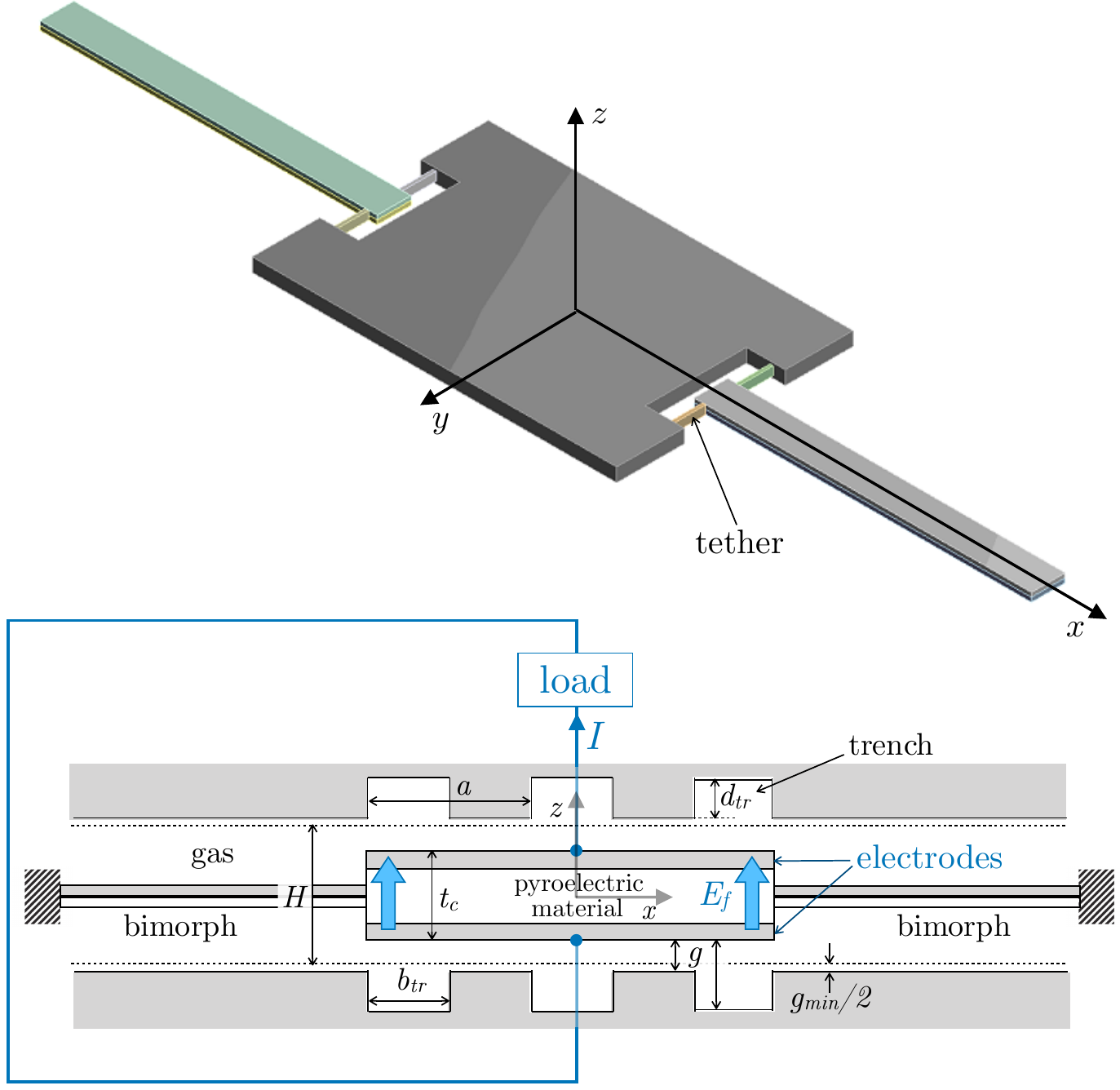}
\caption{The proposed pyroelectric energy harvester. Also shown is connecting the pyroelectric element under the application of an electric field to a load.}
\label{figNewDevice}
\end{figure}

The proposed device, which has a symmetric design, is presented in
Fig. \ref{figNewDevice}. The pyroelectric capacitor, shaped as a rectangular
plate of large area in the $x-y$ plane, is connected to the bimetal cantilevers
by means of four tethers.
Since heat exchange between the pyroelectric capacitor and the heat reservoirs
takes place directly over a large area (in the $x-y$ plane), the time it takes to thermally charge
and discharge the capacitor is very small.
This, in addition to designing the tethers to have sufficiently large thermal
resistance, guarantees a nearly spatially uniform temperature distribution in
the pyroelectric capacitor at all times. The tethers are also designed to have low torsional and bending stiffness which allows them to rotate
about their axis (along the $y$ direction) and bend in the $x$-direction with
negligible effect on the pyroelectric plate so that it remains rigid and horizontal during
the actuation.

Each bimetal cantilever is composed of two metals of different heat expansion
coefficients which force it to bend with temperature variation.
When the pyroelectric capacitor gets in contact with the HTR,
its temperature rises (approaching that of the HTR) and heat diffuses toward the bimetal cantilever
through the tethers. Choosing the upper layer metal to have have the higher heat
expansion coefficient forces the cantilever to bend downward and
push the pyroelectric plate toward the the LTR.
Upon contact with the LTR, the pyroelectric plate
temperature decreases approaching that of the LTR. As the cantilevers cool
down, they bend upward. The cycle starts again when the plate
reaches the HTR.

In addition, the proposed design of Fig. \ref{figNewDevice} does not require direct (solid-solid) contact
between the capacitor and the reservoirs. We assume that the plate stops when it gets
within a distance of $g_{min}/2$ from either reservoir. Although this results in reduction in the
output power and efficiency due to the additional thermal resistance of the gas film (of thickness $g_{min}/2$),
this sacrifice alleviates all the challenges associated with physical contact, and as such increases the life of the device.
Although we do not discuss the stopping mechanism in this work, we point out that it is possible, in theory, to design the device
such that it oscillates in an autonomous fashion without hitting either reservoir and without the need for an additional stopping mechanism.

\section{Thermodynamics Model}
\label{sec:Thermodynamics_Model}
A pyroelectric material is a special type of piezoelectric materials\cite{CambPiezo} that undergoes spontaneous
polarization when its temperature varies in time. 
In engineering applications, the pyroelectric material is typically oriented along the principal crystallographic direction\cite{CambPyro, IOP}. In the presence of an external electric field, $E_f$, applied along the same direction, the pyroelectric effect is quantified in terms of the
generalized pyroelectric coefficient, $\pi=\partial D/\partial T$, measuring the increase in surface the charge density\footnote{The magnitude of the electric displacement at the surface of a conductor.}, $D=\mathcal{E} \; E_f + {\Pi_{\rm{s}}}$, per unit increase in the temperature:
\begin{equation}\label{p_g}
\pi = \pi_s(T) + E_f \frac{{\partial \mathcal{E} }}{{\partial T}}
\end{equation}
where
\begin{equation}\label{pyroscalar}
\pi_s(T)=\frac{\partial \Pi_s}{\partial T} < 0 \ \mbox{for} \ T < T_C,
\end{equation}
$\Pi_s$ is the spontaneous polarization, $T$ is the temperature, $T_C$ is the Curie temperature, and $\mathcal{E}$ is the electrical permittivity of the pyroelectric.

The current ${I_\pi}$ flowing to the external circuit, shown in Fig. \ref{figNewDevice}, is then expressed as:
\begin{equation}\label{i_p}
I_\pi=A\frac{\partial {D}}{\partial {t}}= A \, \pi \frac{dT}{dt}
\end{equation}
where $A$ is the area of the electrode plates. It should be noted that the effect of permittivity variation with temperature can be comparable to the spontaneous pyroelectric effect $\pi_s$ and can also be seen above the Curie point \cite{CambPyro}.

\begin{figure}[!ht]
\centering
\includegraphics[width =5in]{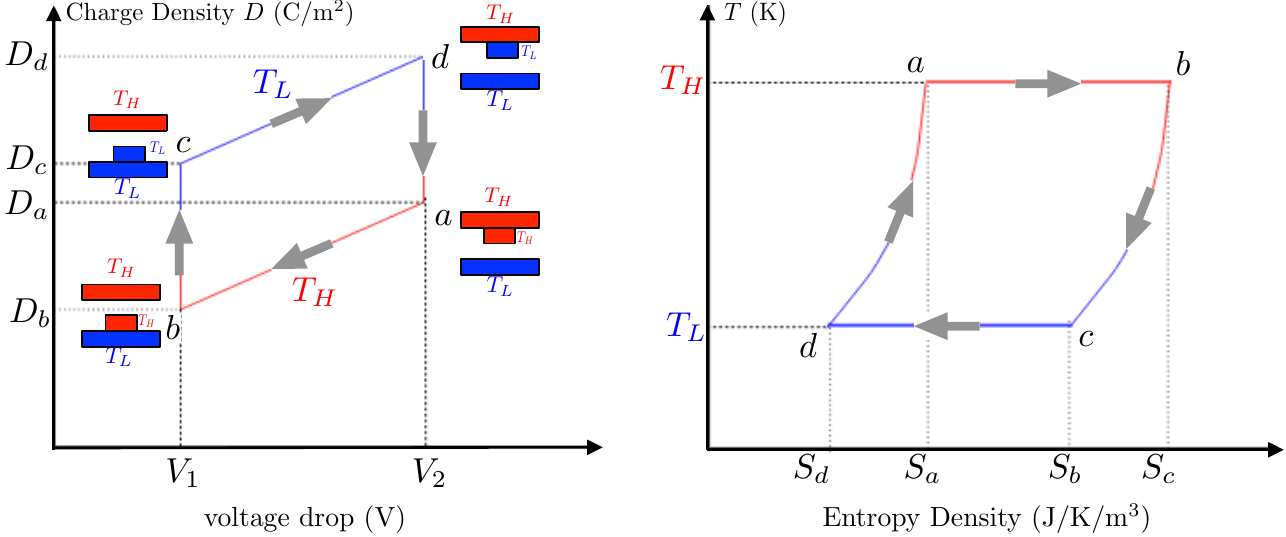}
\caption{Ericsson cycle of the pyroelectric material shown on charge voltage diagram\cite{2012,2011}.}
\label{image:cs1}
\end{figure}

As the plate, shown in Fig. \ref{figNewDevice}, travels back and forth between the HTR and the LTR, the pyroelectric material (sandwiched between the capacitor electrodes), undergoes the Ericsson cycle shown shown in Fig. \ref{image:cs1}. Although the thermodynamic cycle and voltage control are well described in \cite{2012,2011}, \textcolor{red}{we present the cycle here for the sake of completeness and clarity of the observations we made at the end of this section.}
The four processes that make up the cycle are as follows. (1) Process $a \rightarrow b$ is an isothermal process, where the pyroelectric element, of temperature that has just reached $T_H$ by receiving heat from the HTR (state $a$ ), is deflected by the bimetal cantilevers to state $b$, where the pyroelectric element has just been brought in thermal contact with the LTR. During this process, the voltage drops from $\mathcal{V}_2$ to $\mathcal{V}_1$. The voltage is controlled by a sensor as described in Ref. \cite{2011}. The heat gained by the pyroelectric element per unit volume during this isothermal process is $q_{a \rightarrow b}^{\leftarrow}=|\pi| T_H\frac{\mathcal{V}_2-\mathcal{V}_1}{t_c}$, where $t_c$ is the thickness of the pyroelectric element. (2) Process $b \rightarrow c$ is a constant voltage process during which heat is transferred from the pyroelectric element to the LTR. The heat lost per unit volume is $q_{b \rightarrow c}^{\rightarrow}=c_E (T_H - T_L)$, where $c_E$ is specific heat of the pyroelectric element. (3) Process $c \rightarrow d$ is an isothermal process during which the pyroelectric element, at $T_L$, is pulled back toward the HTR by the bimetallic cantilevers as they unbend due to cooling. The heat lost by the pyroelectric element per unit volume during this isothermal process is $q_{c \rightarrow d}^{\rightarrow}=|\pi| T_L\frac{\mathcal{V}_2-\mathcal{V}_1}{t_c}$. (4) Process $d \rightarrow a$ is a constant voltage during which heat is transferred from the HTR to the pyroelectric element. The heat gained per unit volume is $q_{d \rightarrow a}^{\leftarrow}=c_E (T_H - T_L)$. 
With $\frac{\mathcal{V}_2 - \mathcal{V}_1}{t_c} = E_{f2} - E_{f1}$, the net work produced per  cycle is then
\begin{equation}
W^{\rightarrow}_{\mbox{cycle}} = A_c t_c |\pi| (T_H -T_L) (E_{f2} - E_{f1})
\label{EQwork}
\end{equation}
The power production is then the product of $W^{\rightarrow}_{\mbox{cycle}}$ and the cycling frequency $f$,
\begin{equation}
\dot{W}^{\rightarrow} = f A_c t_c |\pi| (T_H -T_L) (E_{f2} - E_{f1})
\label{EQpower}
\end{equation}
The thermodynamic efficiency is
\begin{equation}
\eta=\frac{w^{\rightarrow}}{q^{\leftarrow}_{d \rightarrow a}}= \eta_{C}\frac{1}{1+\beta \eta_{C}}
\label{EQefficiency}
\end{equation}
where the Carnot efficiency is $\eta_C = 1 - \frac{T_L}{T_H}$ and
\begin{equation}\label{eq:beta}
\beta=\frac{c_E t_c}{|\pi|(\mathcal{V}_2-\mathcal{V}_1)}=\frac{c_E}{|\pi|(E_{f2}-E_{f1})}
\end{equation}
Pyroelectric properties of high-quality (0001)AlN films grown on (111) Si were experimentally measured using the dynamic method in Ref. \cite{fuflyigin2000pyroelectric}. The reported value of the pyroelectric coefficient was in the range of 6-8 $\mu C/(m^2 K)$, over a temperature range of 15-35 $\degree C$. With $\mathcal{E}_r=8.5$ at room temperature\cite{akasaki1967infrared}, the corresponding range of the $p/\mathcal{E}_r$ figure-of-merit of 0.8-0.95. The pyroelectric coefficient studied in Ref. \cite{fuflyigin2000pyroelectric} was independent of temperature and applied bias. The specific heat of AlN is $c_E=780$ J/kg $\degree C$ \cite{AlN_Dielectric_Strength}. In Ref. \cite{stan2015electric}, the reported range of values of the pyroelectric coefficient, measured experimentally at room temperature ($298^\circ K$), is 10 - 12 $\mu C/(m^2 K)$ .
Dependence of the relative permittivity on the temperature can be inferred from Ref. \cite{engelmark2003electrical}, where the variation of the capacitance with temperature was studied. Over a temperature range of $25-50$ $\degree C$, $\mathcal{E}_r$ decreases with $T$ as $\frac{d\mathcal{E}_r}{dT} \simeq -1.2147 \times 10^{-3} $ \degree K$^{-1}$. With $\mathcal{E}=\mathcal{E}_r \mathcal{E}_0$, the pyroelectric coefficient, expressed in Eq. (\ref{p_g}), may be approximated as $\pi \simeq \pi_s$, given that the electric field should not exceed the dielectric strength.

Upon inspecting Eqs. (\ref{EQwork}-\ref{eq:beta}), we make the following observations. Assessing the efficiency of the pyroelectric device as a heat engine is dependent on the nature of the HTR as an energy source, given that the LTR is an infinite sink; i.e. it maintains its temperature. If the HTR is an infinite source, i.e. it maintains its temperature at all times as it exchanges heat with the pyroelectric element, then the frequency of operation is more important than how close the efficiency is to the Carnot efficiency. For given $T_H$, $T_L$, and pyroelectric material properties ($\pi$ and $c_E$), the ratio of $\eta/\eta_C$ can be increased by increasing $E_{f2}-E_{f1}$. AlN possesses a relatively high dielectric strength\cite{AlN_Dielectric_Strength} of $E_{f,Break} = 17kV.mm^{-1}$. This property allows applying large electric field difference in the Ericsson's cycle leading to large voltage difference. For $E_{f2}-E_{f1} < E_{f,Break}$, $\beta > \beta^\star$, where $\beta^{\star} = \frac{c_E}{|\pi| E_{f,Break}} = 6.55$. It can be observed from Eq. (\ref{EQefficiency}) that $\eta \rightarrow \eta_C$ for $\eta_C \beta << 1$ (or $\frac{T_L}{T_H} >> 1 - \beta^{\star -1}$) and $\eta \rightarrow \frac{1}{\beta^\star}$ for $\eta_C \beta >> 1$ (or $\frac{T_L}{T_H} << 1 - \beta^{\star -1}$). For $T_L=10$ $\degree C$ and $T_H = 34$ $\degree C$, $\eta_C=0.0781$ and $\eta = 0.66 \eta_C$.
For given $T_H$, $T_L$, pyroelectric material properties ($p$ and $c_E$), and $E_{f2}-E_{f1}$, the power production (Eq. (\ref{EQpower})) can be increased by increasing the operation frequency, $f$, and the volume of the pyroelectric capacitor $A_c t_c$. Note that a smaller thickness of the capacitor implies that one needs to apply a smaller voltage across the capacitor, for given value of the $E_f$. Another key aspect of assessing the performance is whether the power output should be reported per unit area of the device or per unit volume. Reporting the power density as the power output per unit volume may make more sense because it takes into account the distance between the HTR and the LTR. It is also more representative than the surface density when the devices are stacked. The power produced per unit volume is
\begin{equation}
\dot{\omega}^{\rightarrow} = \frac{f L_c t_c |\pi| (T_H -T_L) (E_{f2} - E_{f1})}{(L_c + 2 L_b) H}
\label{EQpowerDensity}
\end{equation}
Note that $T_H$ and $T_L$ are the maximum and minimum average temperature of the pyroelectric capacitor over a cycle.

\section{The Reduced Order Model}\label{sectionReducedOrderModel}
Reduced order modeling\cite{ISSA2014JFE}  is an essential tool to make affordable the cost of computer-based design cycles. In the section, we present the reduced order model of the thermo-mechanical
dynamic operation. The model also accounts for the damping forces imparted on the capacitor by the surrounding
gas. The model is validated by comparing with detailed numerical simulations using Ansys, as presented in Section{modelValidation}.  Being two orders of magnitude faster, the reduced order model is used to optimize the design over wide ranges of design parameters.
To this end, the model is embedded within an optimization algorithm to estimate the dimensions that maximize the power produced per unit volume
for a given temperature difference between the HTR and the LTR.

\begin{figure}[!ht]
\centering
\includegraphics[width = 4in]{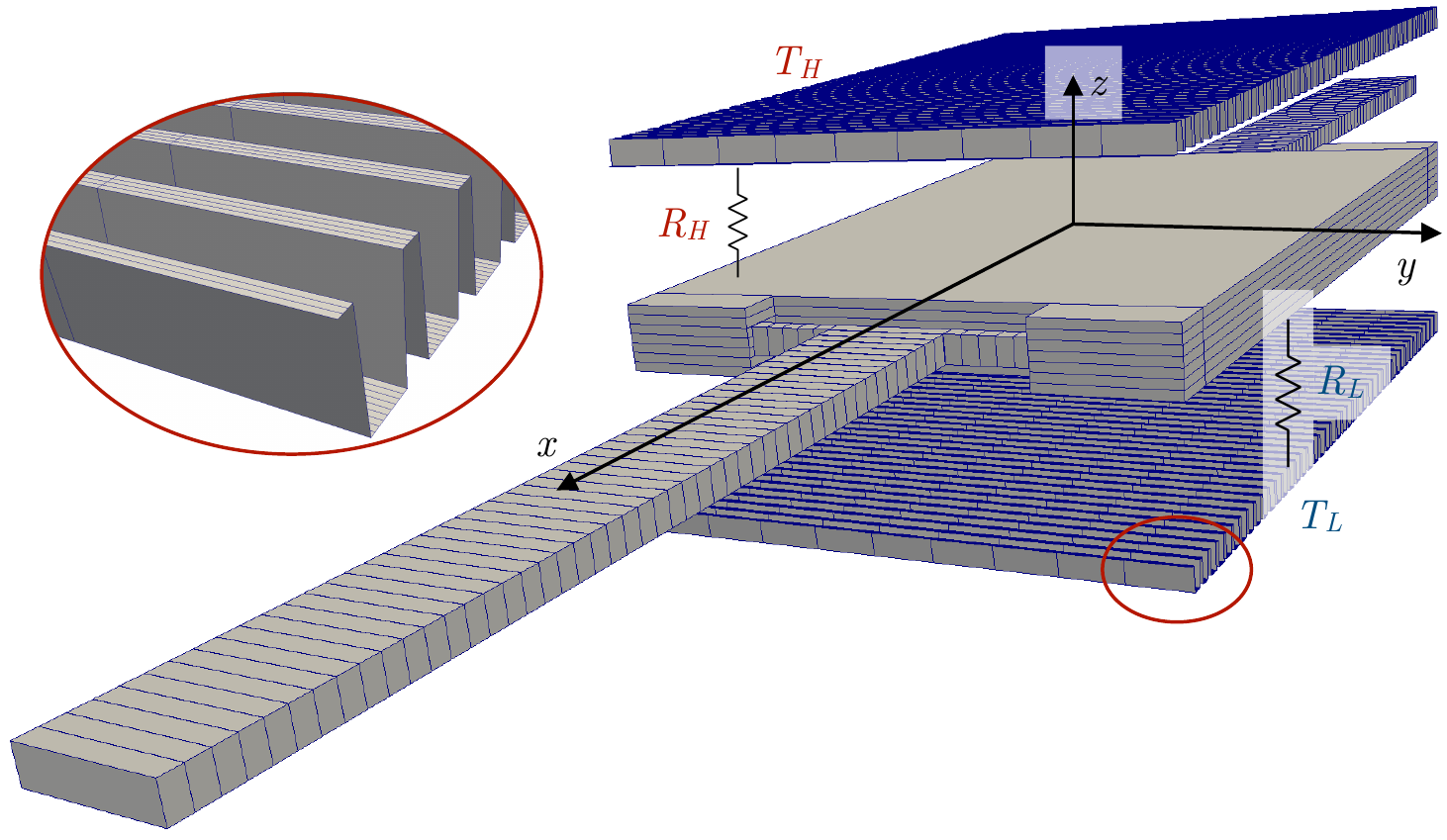}
\caption{Representative meshes used in the reduced order model for the heat transfer (capacitor, tethers, and bimorphs) and the squeeze film (trenched thermal reservoirs).
}
\label{image:ReducedOrderModelMesh}
\end{figure}

The motion of the pyroelectric capacitor, assumed to be rigid of mass $m_c$ and position $z_c$,
is governed by
\begin{equation}
	m_c \frac{d^2 z_c}{dt^2} = F_{tip}(\Delta T, z_{tip}) + F_{torsion} + F_{damping}
	\label{eq:Newton}
\end{equation}
where $F_{tip}(\Delta T, z_{tip})$ is the thermo-mechanical force imparted by the bimetal cantilevers,
$F_{torsion}$ is the torsion force imparted by the tethers, and $F_{damping}$ is the damping force
imparted by the surrounding gas. The thermo-mechanical force imparted by the bimetal cantilevers on the capacitor
takes into account the deformation due to thermal expansion (contraction) caused by temperature changes.
Next, we present models for calculating these forces along with the temperature distribution in the device. Numerical solution 
of the model equations require spatial discretization of the different elements, as depicted in Fig. \ref{image:ReducedOrderModelMesh}.

\label{sec:SolidMech}
\subsection{Forces by the bimetal cantilever and the tethers}
In this section, we present an expression for the force exerted by the bimetal cantilevers on the pyroelectric capacitor. We choose the tethers to be sufficiently compliant so that the horizontal force is small compared to the vertical force. This requires the bending stiffness of the tethers in the horizontal $x$-direction to be  small compared to the bending stiffness in the vertical direction $z$-direction, i.e. $(b/t)_t << 1$. The distance between the tethers and the pyroelectric capacitor must be sufficiently large to provide the needed room for bending.
\begin{figure}[!ht]
\centering
\includegraphics[width=2.75in]{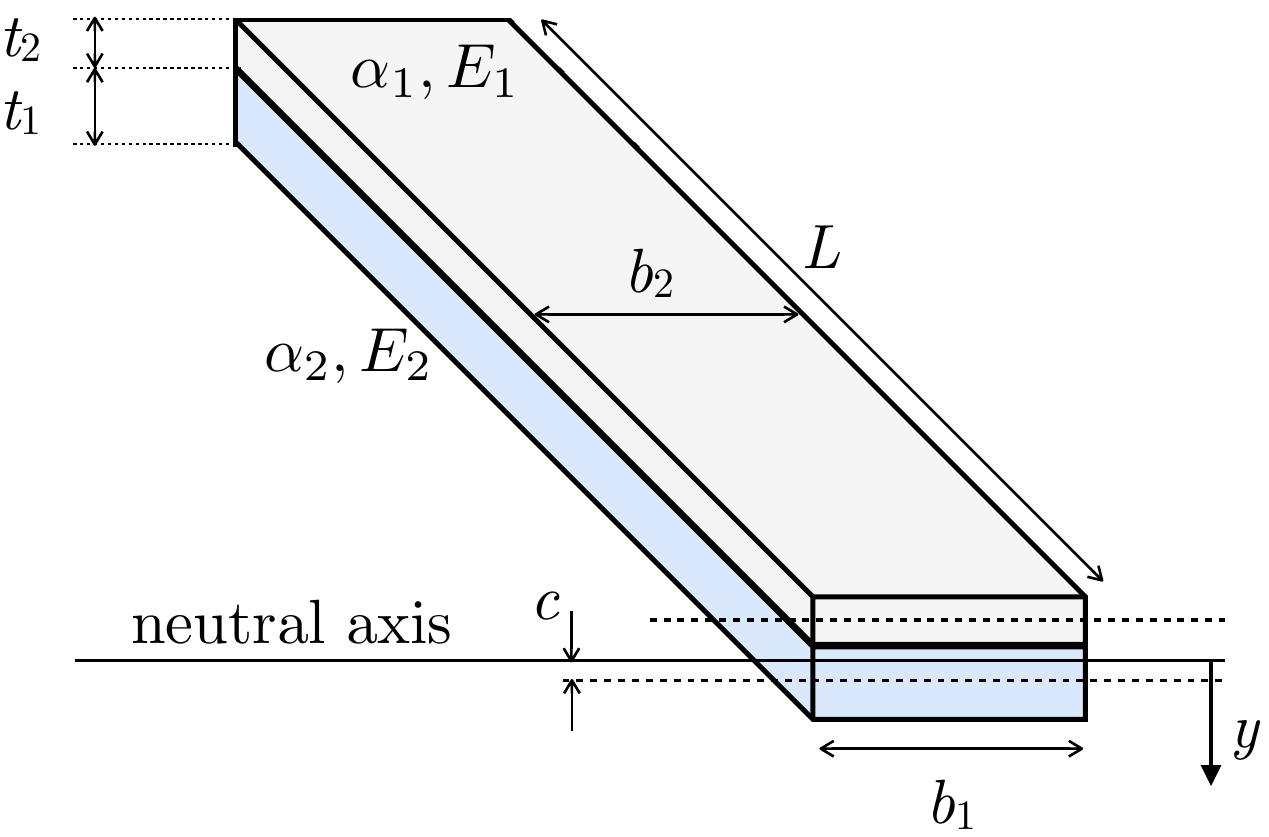}
\caption{Schematic of the bimetal cantilever.} 
\label{fig:bimetal cantilever1}
\end{figure}
The vertical force exerted by the beams on the pyroelectric capacitor is a function of their temperatures and tip deflections. The (downward) force at the tip of the bimetal cantilevers is the sum of a linear spring component and a thermal expansion component
\begin{equation}
F_{tip}(\Delta T,z_{tip})=\left( z_{tip}-z_{tip,exp} \right) \frac{3(EI)_{eq}}{L_b^3}
\label{eq_ftipbimorpth}
\end{equation}
where $z_{tip,exp}$ is the tip deflection induced by thermal expansion. For a spatially uniform temperature distribution $z_{tip,exp}=\frac{K L_b^2}{2}$, 
where the curvature, $K$,  of the bimetal cantilever due to thermal expansion can be found in Ref.\cite{tipdef}
\begin{equation}
K=\frac{6 b_1 b_2 E_1 E_2 t_1 t_2 (t_1 +t_2 )(\gamma _2 -\gamma_1 )\Delta T}{(b_1 E_1 t_1 ^2)^2+(b_2 E_2 t_2 ^2)^2+2b_1 b_2 E_1 E_2 t_1 t_2 (2t_1^2+3t_1t_2+2t_2^2)},
\label{eq:Curvature}
\end{equation}
$(EI)_{eq}$ is the equivalent flexural rigidity and $\Delta T$ is the change in temperature of the bimetal cantilevers from a reference value, $T_0$, for which $K_0 = 0$. The various dimensions in Eq. (\ref{eq:Curvature}) are depicted in Fig. \ref{fig:bimetal cantilever1}. The equivalent flexural rigidity, $(EI)_{eq}$, is 
\begin{equation}
(EI)_{eq}=E_1\left[\frac{b_1t_1^3}{12}+\left(\frac{t_1+t_2}{2}-c \right)^2 b_1t_1\right]+ E_2\left[\frac{b_2t_2^3}{12}+c^2 b_2t_2\right]
\end{equation}
where $c$, the location of the common neutral axis relative to the axis of the lower beam, is
\begin{equation}
c=\frac{E_2b_2t_2t_1}{E_2b_2t_2+E_1b_1t_1}
\label{eq_commonNeutralAxis}
\end{equation}
Since the heat transfer in the bimetal cantilever is slow compared to that in the pyroelectric capacitor, the temperature distribution along the bimetal cantilever is non-uniform. In this case, the beam is discretized into a number elements each of which assumes a uniform temperature. The tip deflection, arising from thermal expansion, is obtained as
\begin{equation}
z_{tip,exp}=\frac{L_b^2}{n^2}\bigg[\frac{K_n}{2}+\frac{3}{2}\sum_{i=1}^{n-1}K_i +\sum_{i=1}^{n-2}\sum_{j=1}^{i}K_j\bigg],\ \ n \geq 3
\label{eq:sol_recursive}
\end{equation}
where the curvature $K_i$ of element $i$ calculated from Eq. (\ref{eq:Curvature}).

Since the bimetal cantilever is connected to the plate through the tethers, it is necessary to factor in the impact of the tethers on the force at the tip of the bimetal cantilever. The contribution of the tethers to this force can be divided into two independent components: torsion and bending of the tethers. 
Torsion of the tethers occurs due to the bending of the bimetal cantilever while the pyroelectric plate stays horizontal (due to symmetry) as it shuttles back and forth between the two thermal reservoirs. At the same time, the tethers experience bending in the $x-$direction caused by the thermal expansion/contraction of the bimetal cantilever.  Since the tethers are designed so that this bending compliance is large, we neglect the associated horizontal force component on the bimetal cantilever tip, as discussed at the beginning of this section. The tethers torque, $\mathcal{T}$, is transformed to a bending moment on the tip of the bimetal cantilever element. The equivalent vertical force at the tip of the bimetal cantilever that yields the same tip displacement as that due to the moment, assuming small deflections, is
\begin{equation}
F_\textrm{torsion}=\frac{3}{2}\frac{\mathcal{T}}{L_\textrm{b}}, \mathcal{T} =  \left(\frac{JG}{L}\right)_t \theta
\end{equation}
where $G$ is the shear modulus, $L_t$ is the length the tether, $\theta$ is the angle of twist at the tip of the tether and $J \simeq t b^3 \left( \frac{1}{3} - 0.21 \frac{b}{t} \left( 1 - \frac{b^4}{12 t^4} \right)\right)$ (Ref. \cite{young2002roark}).

\subsection{Gas Damping}\label{sectionGasDamping}
In order to realize the desired damping force for optimal performance of the device, the HTR and the LTR are patterned with trenches, as shown in Fig. \ref{figNewDevice}. In arriving at the optimized designs presented in Section \ref{optimalDesigns}, the trench width and depth are adjusted so that the pyroelectric plate is sufficiently damped while maximizing the energy produced. 


The motion of the pyroelectric capacitor is such that it remains parallel to the thermal reservoirs. In addition, the gap ($g$) separating the capacitor and each of the reservoirs is much smaller than the plate length ($L_c$) and width ($b_c$). At an operation frequency of $f= 60 Hz$, the average plate speed is $V_c = 2 H f = 0.0024 m/s$. The average speed at which the air is pushed as the plate squeezes the air film is then $V = V_c L_c b_c / (\Pi_c H) = 0.0074 m/s$, where $\Pi_c = 2(L_c + b_c)$ is the plate perimeter. The Reynolds number of the flow is $\mathtt{Re}=V H/\mu' = 0.01$, where $\mu'=1.48 \times 10^{-5} m^2/s$ is the kinematic viscosity of air at atmospheric conditions.
Since $\mathtt{Re}<<1$, the flow can be modeled as inertia free. In addition, the flow is assumed to be isothermal. This assumption is justified on the grounds that viscous dissipation does not lead to any appreciable temperature rise and that, in non-insulated micro-devices, the heat generated will be lost quickly due to the large surface to volume ratio. The fact that the squeeze film thickness (the gap separating the pyroelectric plate from either reservoir) is much larger than the mean free path, $\lambda$, of air (at the thermodynamic state corresponding to the enclosure pressure and ambient temperature) implies that the flow is in the continuum regime, i.e. $\mathtt{Kn} = \lambda/H <<1$. The only exception is when the pyroelectric plate gets too close (a few mean free paths) to either reservoir.
Under the assumptions of isothermal and inertia-free flow in the continuum regime, we can use the lubrication theory whereby the pressure,
$p(x,y,t)$, of the air film between the pyroelectric plate and either reservoir is governed by the Reynolds equation
\begin{equation}
\frac{\partial \left(p g \right) }{\partial t} = \frac{1}{12 \mu'}  \left[ \frac{\partial}{ \partial x} \left(p g^3  \frac{\partial p}{\partial x}  \right) +  \frac{\partial}{\partial y} \left( p g^3\frac{\partial p}{\partial y}  \right) \right]
\label{eqDampingPDE1}
\end{equation}
where $g(x,y,t)$ is the gap separating the pyroelectric plate and the thermal reservoir (see Fig. \ref{figure:pyroWithTrenches}), and $\mu'$ is the effective viscosity \cite{veijola2002end,veijola1995equivalent}, expressed as
$\mu'=\frac{\mu}{1+9.638 \, \mathtt{Kn}^{1.159}}$. 

We assume that the change in pressure, $\hat{p}$,  introduced by the motion of the pyroelectric capacitor is a small compared to the unperturbed enclosure pressure, $p_0$. Then for $p=p_0 + \hat{p}, \,\,\, \hat{p} << p_0$, Eq. (\ref{eqDampingPDE1}) is approximated as
\begin{equation}
\frac{\partial g }{\partial t} = \frac{1}{12 \mu}  \left[ \frac{\partial}{ \partial x} \left(g^3  \frac{\partial \hat{p}}{\partial x}  \right) +  \frac{\partial}{\partial y} \left( g^3\frac{\partial \hat{p}}{\partial y}  \right) \right]
\label{eqDampingPDE2}
\end{equation}
The damping force acting on the plate is then computed by integrating, over the area of the pyroelectric capacitor area, the difference in pressure between the bottom and top surfaces  
\begin{equation}
F_{\mbox{damping}} = \int_{W_p} \int_{L_p} \left( \hat{p}_b(x,y,t) - \hat{p}_t(x,y,t) \right) \, dx dy
\label{eq:dampingForce}
\end{equation}

Since pyroelectric plate travels the entire distance separating the two reservoirs, solutions based on the linearized Reynolds equation\cite{blech1983isothermal} for small displacements are not applicable here. The nonlinear partial differential Eq. (\ref{eqDampingPDE2}) is numerically solved in each of the two gas film domains between the capacitor and the two thermal reservoirs. The numerical method uses an explicit first order
time integration scheme with second order finite difference discretization of the spatial derivatives. Figure \ref{image:ReducedOrderModelMesh} shows an example of the mesh used to represent the surface of the trenched thermal reservoirs when solving Eq. (\ref{eqDampingPDE2}) .

\label{sec:heatTransfer}
\subsection*{Heat Transfer}
The heat model is based on dividing the device into three different zones as depicted in Fig. \ref{image:ReducedOrderModelMesh}: the pyroelectric plate, the tethers, and the bimetal cantilever. The model also takes into account the thermal resistance of the gas film between the capacitor and the thermal reservoirs.  Each zone is modeled by the discretized 1D heat equation (along the dominant component of the heat flux) using second order central difference scheme for the diffusion term and first order explicit time integration scheme, as presented next for the capacitor. Similar discretizations were carried out for the tethers and bimetal cantilevers in the $y$ and $x$ directions respectively.


The pyroelectric capacitor temperature, $T_c(z,t)$, is assumed to be uniform in the ($x-y$) plane of the capacitor. Its variation along the $z$ direction is captured by discretizing the capacitor into layers in the $x-y$ plane, where the energy balance for layer $i$ of thickness $\Delta z$ is governed by
\begin{equation}
\rho_c c_c \Delta V_i \frac{T_c(z_i,t+\Delta t)-T_c(z_i,t)}{\Delta t}=\dot{Q}_i^{\leftarrow} - \dot{Q}_i^{\rightarrow}
\end{equation}
where $\Delta V_i = L_c b_c \Delta z$. For the interior nodes, the rates of heat transfer into and out of the layer are
\begin{eqnarray}
& \dot{Q}_i^{\leftarrow} &= k_c A_c \frac{T_c(z_{i+1},t)-T_c(z_i,t)}{\Delta z_c}, \nonumber \\
& \dot{Q}_i^{\rightarrow} &= k_c A_c \frac{T_c(z_{i},t)-T_c(z_i-1,t)}{\Delta z_c} + \delta_{ii^*} \frac{T_c(z_{i},t)-T_t(1,t)}{R_{ii^*}} \nonumber
\end{eqnarray}
where $A_c=L_c b_c$, $A_t = b_t t_t$, $i^*$ refers to capacitor element $i$ in contact with tether with intersection area $A_{ii^*}$, $\delta$ is the Kronecker's Delta,
and the thermal resistance $R_{ii^*} = \frac{\Delta y_t/2}{k_t (4 A_{ii^*})} + \frac{t_c/2}{k_c (4 t_c^2 (A_{ii^*}/A_t))}$.
For $i=N_p$,
\begin{equation}
\dot{Q}_i^{\leftarrow} = \frac{T_H-T_c(N_p,t)}{R_H+ \frac{\Delta z_c/2}{k_c A_c} }, \nonumber
\end{equation}
and for $i=1$,
\begin{equation}
\dot{Q}_i^{\rightarrow} = \frac{T_c(1,t)-T_L}{R_L+ \frac{\Delta z_c/2}{k_c A_c}}
\end{equation}
where $R_H$ and $R_L$ are the thermal resistances of the gas film between the capacitor and the high and low temperature reservoirs respectively. Referring to
Fig. \ref{figNewDevice}, these resistance are approximated as
\begin{eqnarray}
R_H & = &\frac{min(g)}{k_g (1-\chi) A_c} \parallel \frac{max(g)}{k_g \chi A_c} \\
R_L & = & \frac{min(g')}{k_g (1-\chi) A_c} \parallel \frac{max(g')}{k_g \chi A_c}
\end{eqnarray}
where $\chi$ is the fraction of the capacitor area that is exposed to trenches and $g' = H - g$.

\section{Dynamic Operation}
\label{sec:dynamicOperation}

In this section, we investigate the dynamic behavior of the device using Ansys simulations.
Understanding the dynamic operation of the device is essential for identifying the key design considerations which will in turn enable tightening the ranges of design parameters used in the design optimization procedure presented in Section \ref{sec:designConsiderationsAndOptimization} . The properties of the materials comprising the bimetal cantilever layers, tethers, and pyroelectric capacitor for all the designs considered in this work are presented in Table \ref{table:materialProperties}. Note that these materials are commonly used in MEMS devices \cite{2012}. To aid the discussion in this and the following section, we employ a device with dimensions listed in Table \ref{table:dimensions}.

\begin{table}[!h]
\centering
\begin{tabular}{lcccc}
\toprule
& {cantilever} & {cantilever} & {Tethers}      & {Pyroelectric} \\
& {Layer 1} & {Layer 2} &       & capacitor \\
\midrule
{Material}                                                & Al & Si & Al             & AlN                \\
{$\rho$ ($kg.m^{-3}$)}                        & 2770                     & 2330                     & 2770                 & 3260                            \\
{$c_p$ ($J.kg^{-1}.K^{-1}$)}   & 875                      & 712                      & 875                  & 740                             \\
{$k$ ($W.m^{-1}.K^{-1}$)} & 150                      & 148                      & 150                  & 140                             \\
{$E$ ($GPa$)}                           & 71                       & 190                      & 71                   & 350                             \\
{$\nu$}                                   & 0.33                     & 0.22                     & 0.33                 & 0.24                            \\
{$\gamma$}                  & $2.30\times 10^{-5}$     & $2.33\times 10^{-6}$     & $2.30\times 10^{-5}$ & $4.50\times 10^{-6}$      \\
\bottomrule
\end{tabular}
\caption{Materials (and their properties) selected for the bimetal cantilever layers, tethers, and pyroelectric capacitor.}
\label{table:materialProperties}
\end{table}

\begin{table}[!h]
\centering
\begin{tabular}{l|cccc}
\toprule
& {bimetal cantilever} & {bimetal cantilever} & {Tethers}      & {Pyroelectric} \\
& {Layer 1} & {Layer 2} &     & {Capacitor} \\
\midrule
Length $L$                                 & 1000                     & 1000                     & 20                   & 1000                            \\
Width $b$                                    & 20                       & 20                       & 2                    & 140                              \\
Height $t$                                     & 2.5                      & 2.5                      & 5                    & 8                            \\
\bottomrule
\end{tabular}
\caption{Device dimensions ($\mu m$) used for discussing device operation, design consideration, and for comparing the model to Ansys FSI simulations.}
\label{table:dimensions}
\end{table}

\begin{figure}[!ht]
\centering
\includegraphics[width = 6in]{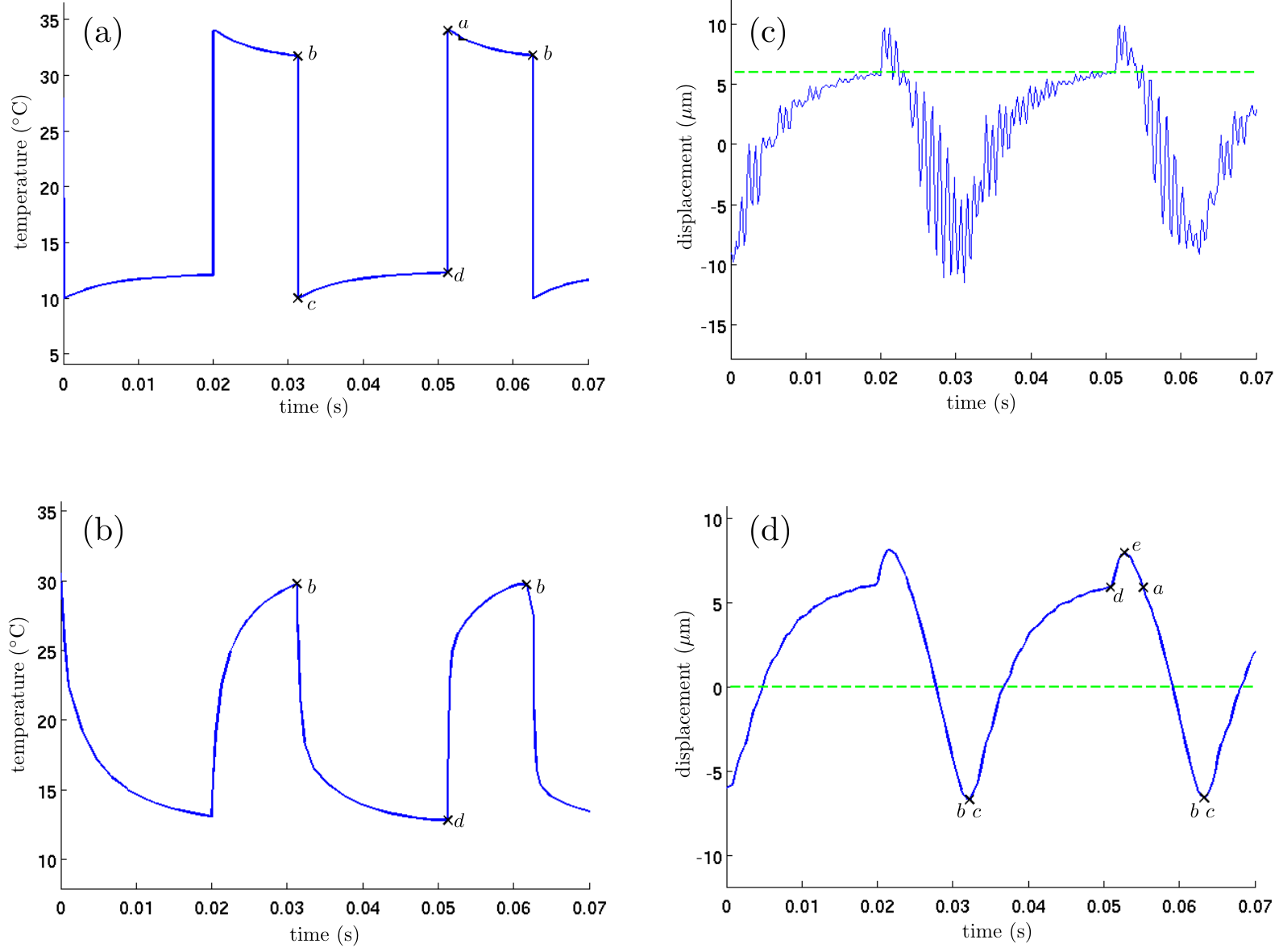}
\caption{Ansys simulation results of (a) the temperature variation at the center of the capacitor, (b) the temperature variation at a point of a bimetal cantilever $200\mu m$ away from the capacitor,
the capacitor center  displacement while the device is operating with the absence (c) and presence (d) of damping.
}
\label{fig:tempSim}
\end{figure}

The gap separating the reservoirs and  the temperature difference between the two reservoirs are selected to be $H = 20\mu m$  and $T_H - T_L = 24^{\circ} C$. For the Ansys simulations of the thermo-mechanical dynamical behavior, thermal contact with the reservoirs is modeled by setting the temperature of the pyroelectric capacitor to that of the reservoir during contact. The alternating temperature boundary conditions are assigned as a function of time depending on the state of contact with the reservoirs. As the capacitor travels between the two reservoirs, we assume that the heat losses to the surroundings are negligible. 
Every time step, the Ansys transient heat transfer solver computes the temperature variation within the device, as shown in Figs. \ref{fig:tempSim}(a) and (b) at two different locations. The temperature distribution is then imported by the transient structural solver which simulates the device motion and the bimetal cantilevers actuation as shown in Figs. \ref{fig:tempSim}(c) and (d).

\subsection{Heat Transfer Dynamics}
Figs. \ref{fig:tempSim}(a), \ref{fig:tempSim}(c) and \ref{fig:tempSim}(d) show the time evolution of the temperature and the displacement of the center of the capacitor. Fig. \ref{fig:tempSim}(b) shows the time evolution of the temperature at a point on the bimetal cantilever located at a distance of 200 $\mu m $ from the capacitor. State $b$ corresponds to the time instant of contact between the capacitor and the LTR where the contacting capacitor surface is assigned the temperature of the LTR ($T_L=10^{\circ}C$). Upon contact, the capacitor cools rapidly ($b \rightarrow c$) and reaches the temperature of the reservoir in almost $5\mu s$. The speed of this heat transfer is in accordance with that predicted by the time scale analysis, Eq. (\ref{eq:tau_Cap}). At state $c$, the capacitor detaches from the reservoir as a result of the pull force induced by the cooling down of the bimetal cantilevers. Once it detaches, the capacitor starts absorbing heat from the bimetal cantilevers 
which explains the capacitor temperature increase over the period $c \rightarrow d$ in Fig. \ref{fig:tempSim}(a) and the corresponding bimetal cantilevers temperature decrease in Fig. \ref{fig:tempSim}(b). At state $d$, the capacitor contacts the HTR. The reservoir temperature then rises quickly ($d \rightarrow a$) to that of the HTR. Once the temperature of the bimetal cantilevers increase sufficiently, the capacitor  is pulled from the HTR (state $a$ in Fig. \ref{fig:tempSim}(d) ) and moves until it contacts the LTR at state $b$. This completes the cycle.


\subsection{Thermo-mechanical Behavior of the bimetal cantilever}
In order to further understand the thermo-mechanical behavior of the bimetal cantilevers underpinning the actuation mechanism, simulations were carried out without imposing any mechanical boundary condition on the displacement of the capacitor at contact with the thermal reservoirs. Upon contact with the HTR at state $d$ (Fig. \ref{fig:tempSim}(d)), the elevation of the capacitor increases toward $e$, indicating that the thermal inertia of the bimetal cantilevers and the capacitor kept pushing the capacitor through the HTR until state $a$ is reached, at which point the capacitor is pulled away from the HTR toward the LTR.
This behavior, observed when the capacitor reaches the HTR, is a consequence of its rapid lateral thermal expansion as it receives heat from the reservoir. This deformation imparts a moment on the bimetal cantilevers that causes them to deflect upwards between points $d$ and $e$ (Fig. \ref{fig:tempSim}(d)).
As heat continues to diffuse into the bimetal cantilevers, bending starts to take effect.
Consequently, the bimetal cantilevers start deflecting downward once state $e$ is reached (Fig. \ref{fig:tempSim}(d)). In contrast, at the impact with the LTR, the lateral thermal contraction of the pyroelectric capacitor upon cooling imparts a moment that also deflects the bimetal cantilevers upward. This effect adds to the upward bending of the bimetal cantilevers as they lose heat to the capacitor. Hence, the capacitor does not fall below the LTR and the duration separating states $b$ and $c$ is small.

Another characteristic of the observed asymmetric behavior is that the capacitor takes more time to travel from the LTR to the HTR  than from the HTR to the LTR. This behavior is attributed to
the component of the vertical force  that arises from expansion/contraction of the pyroelectric capacitor and the bimetal cantilevers. If the tethers are sufficiently stiff in the $x-$direction, then this force will add to the upward force during process $c-d$ and will oppose the downward force during process $a-b$, as depicted in Fig. \ref{fig:tempSim}(d).


\begin{figure}[!ht]
\centering
\includegraphics[width = 4in]{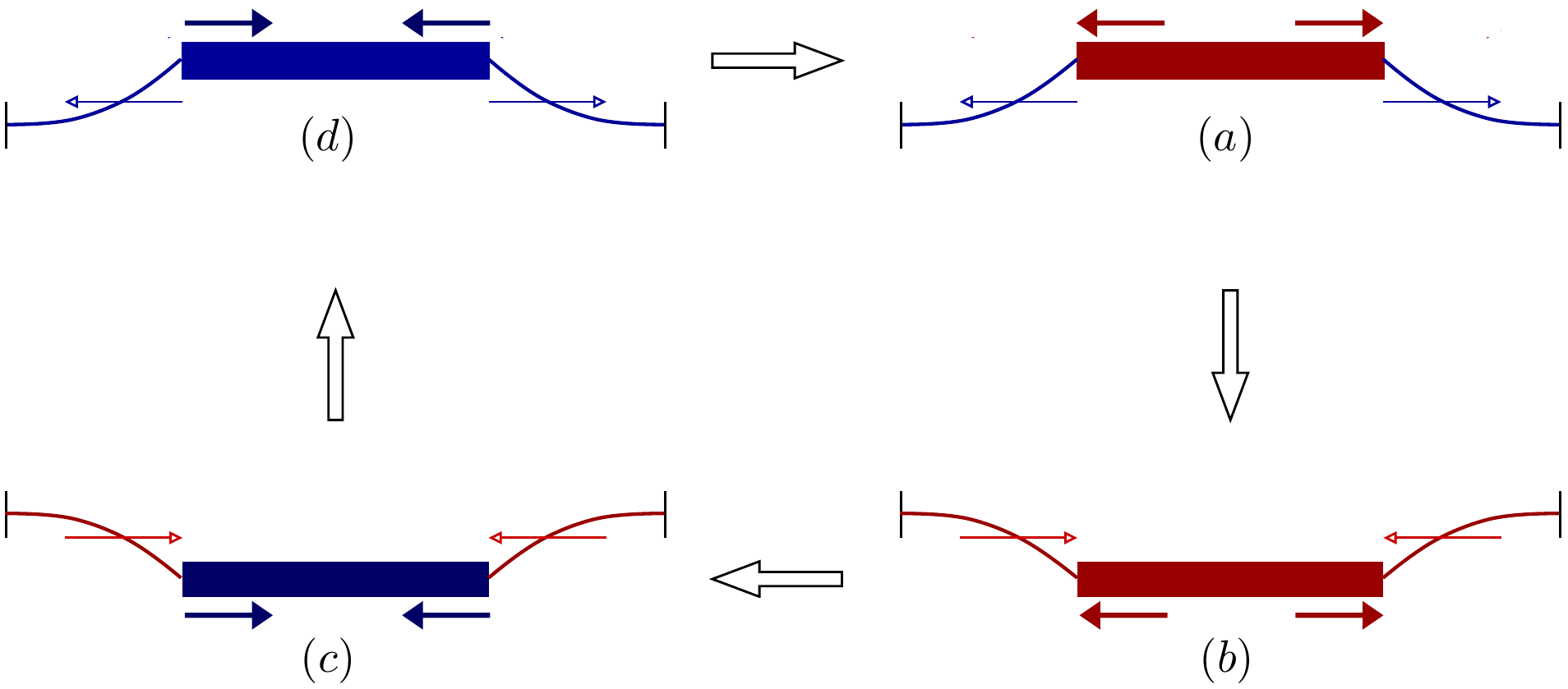}
\caption{Schematic representing, at different states of the cycle, the effect of the thermal expansion/contraction of the capacitor and bimetal cantilevers on the vertical force acting on the capacitor. The resulting moment imparted on the bimetal cantilevers adds to the downward force when the capacitor is traveling from the HTR to the LTR and reduces the upward force when the capacitor is traveling from the LTR to the HTR. This renders the process $c-d$ slower than the process $a-b$.
}
\label{fig:Cycle_Schematic}
\end{figure}

\subsection{Natural Structural Oscillations and Damping}
To observe the natural structural oscillations and the impact of damping, two simulations were carried out, one without damping (Fig. \ref{fig:tempSim}(c)) and one with damping  (Fig. \ref{fig:tempSim}(d)). The two simulations exhibit oscillations between the reservoirs with the same operational frequency of about $30Hz$. The undamped response, shown in Fig. \ref{fig:tempSim}(c), exhibits high frequency oscillations about the damped response of Fig. \ref{fig:tempSim}(d). The frequency of these oscillations is about $1120kHz$, which is within $10\%$ of  the frequency ($1265kHz$)  of the highest energy mode of vertical translational oscillations predicted by modal analysis. When a damping coefficient of  $b=4\times10^{-4}\mu N.s.\mu m^{-1}$ is introduced, these high frequency structural oscillations disappear, as seen in Fig. \ref{fig:tempSim}(d). 

\subsection{Thermo-mechanical Dynamics of the Damped Constrained Oscillations}
As discussed earlier, the $x-$bending stiffness of the tethers cause the operation cycle to be asymmetric, where the capacitor spends more time traveling from the LTR to the HTR than from the HTR to the LTR.
The operation frequency in this case is $\sim 30 Hz$. Increasing the operation frequency is a key design objective, since the power output is the product of the work produced per cycle (Eq. (\ref{EQwork})) and the operation frequency.
To increase the operating frequency, the design is modified so that the asymmetry between the durations $b-c-d$ and $d-a-b$ of Fig. \ref{fig:tempSim}(c) is significantly reduced.
This is realized by increasing the $x-$bending compliance of the tethers by a factor of 10 without affecting the heat transfer dynamics through the tethers.  This may be accomplished by reducing the tether width by a factor of $\sqrt{10}$ and increasing its height by the same factor. 
Reducing the $x-$bending stiffness of the tethers will diminish the contribution of the thermal expansions/contractions of the capacitor/tethers to the vertical force on the capacitor. Instead, these expansions/contractions will contribute to tethers deflections.

\begin{figure}[!ht]
\centering
\includegraphics[width = 4in]{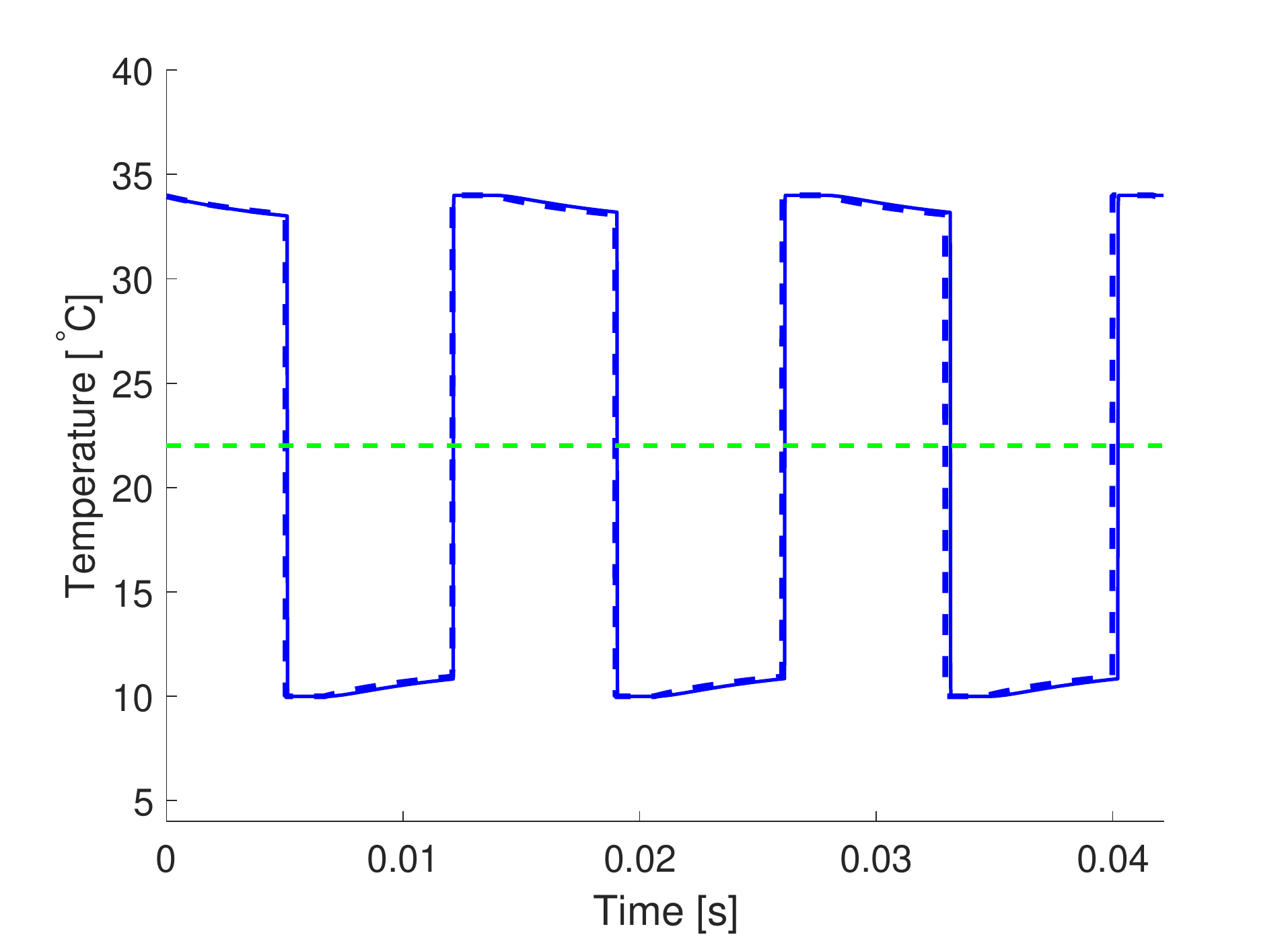}
\caption{Simulation results of the temperature at the center of the capacitor with constrained motion and presence of damping. Reduced-order model: dashed lines. Ansys: solid lines.}
\label{fig:Temp_LowStiff}
\end{figure}

\begin{figure}[!ht]
\centering
\includegraphics[width = 4in]{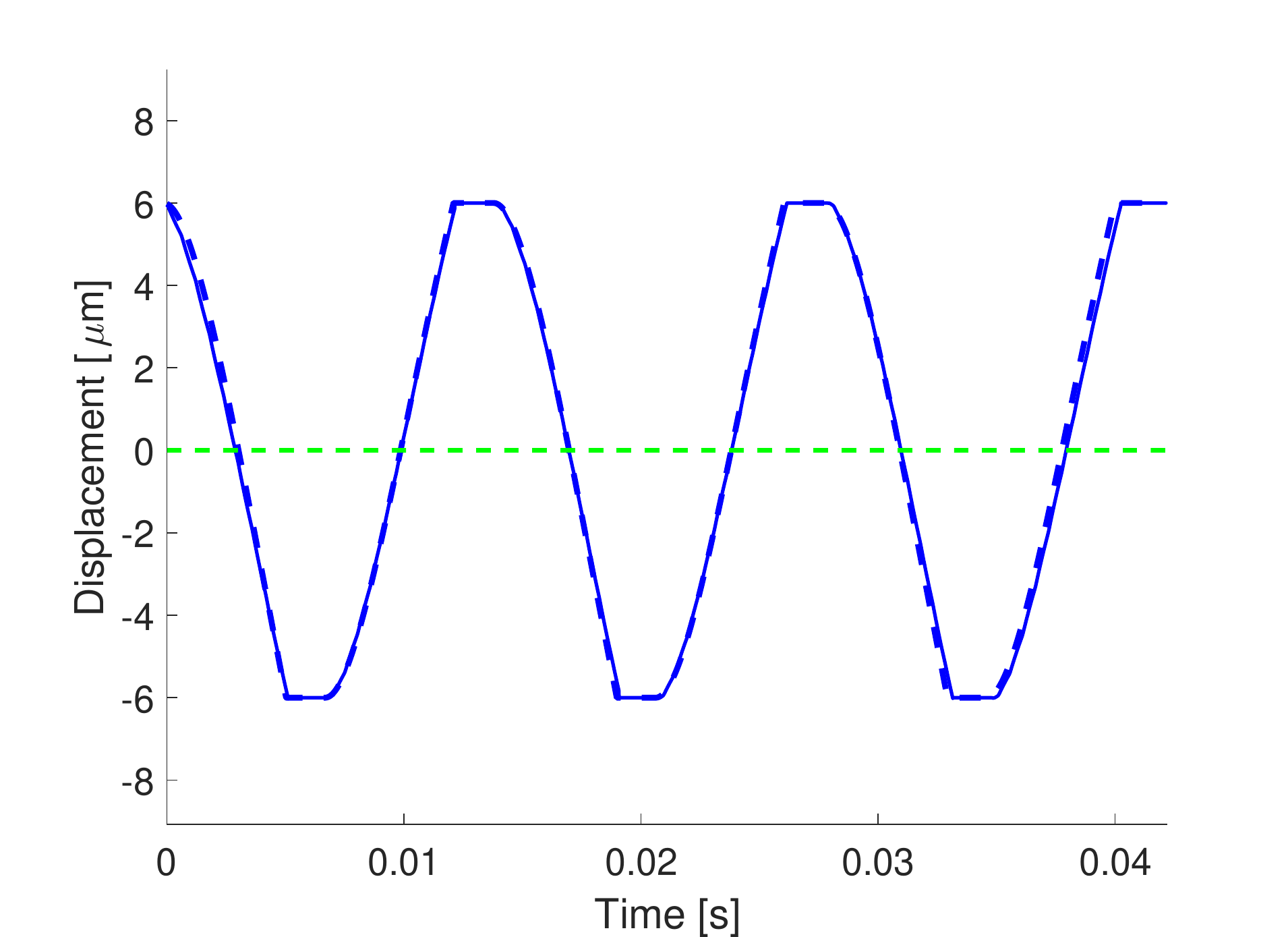}
\caption{Simulation results of  the displacement at the center of the capacitor with constrained motion and presence of damping. Reduced-order model: dashed lines. Ansys: solid lines.}
\label{fig:Pos_LowStiff}
\end{figure}

\begin{figure}[!ht]
\centering
\includegraphics[width = \textwidth]{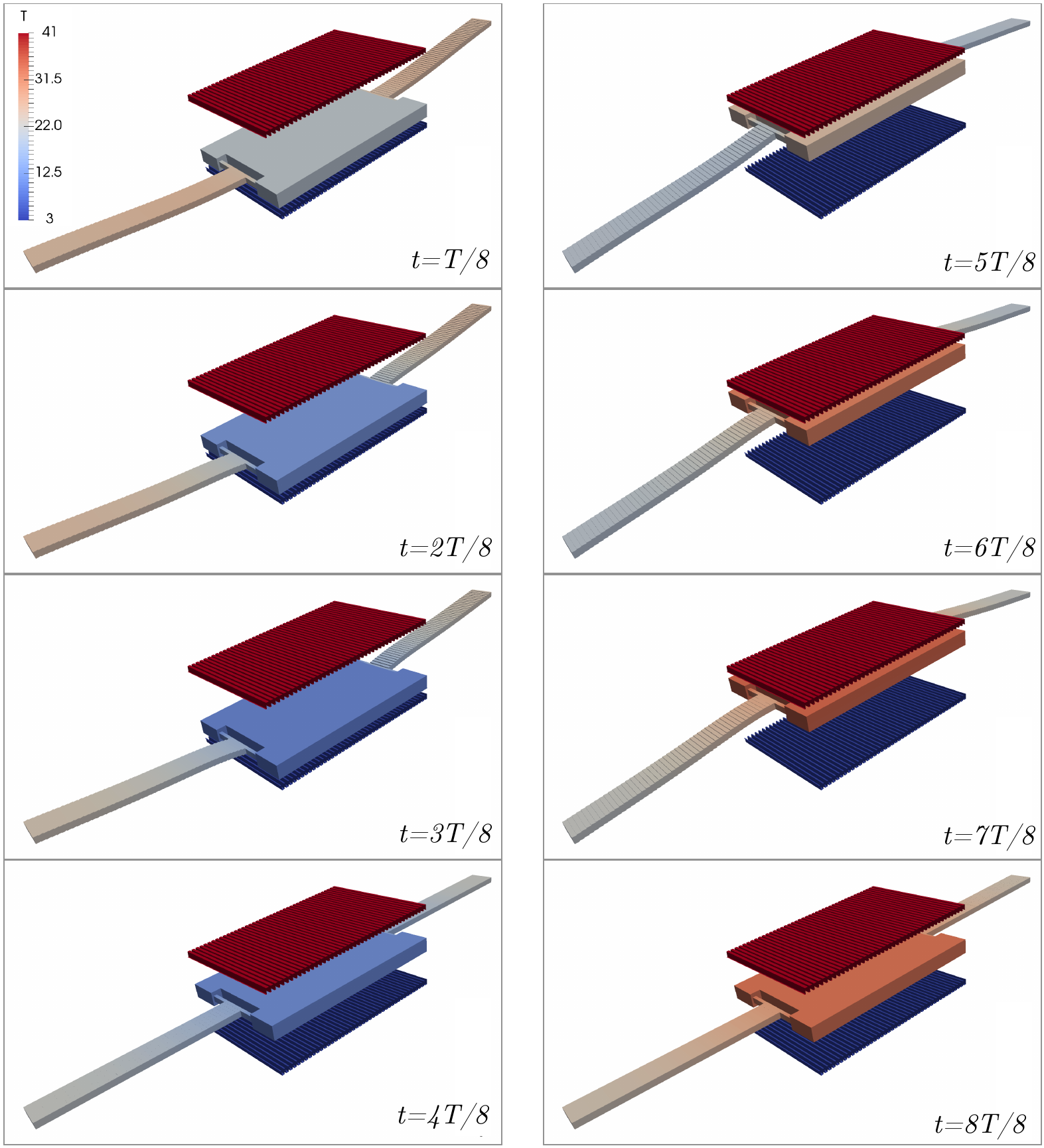}
\caption{Simulation results for case \textcolor{red}{WHAT?} over a complete cycle at the quasi-stationary state. The color denotes the temperature.}
\label{fig:seqT}
\end{figure}

\begin{figure}[!ht]
\centering
\includegraphics[width = 5in]{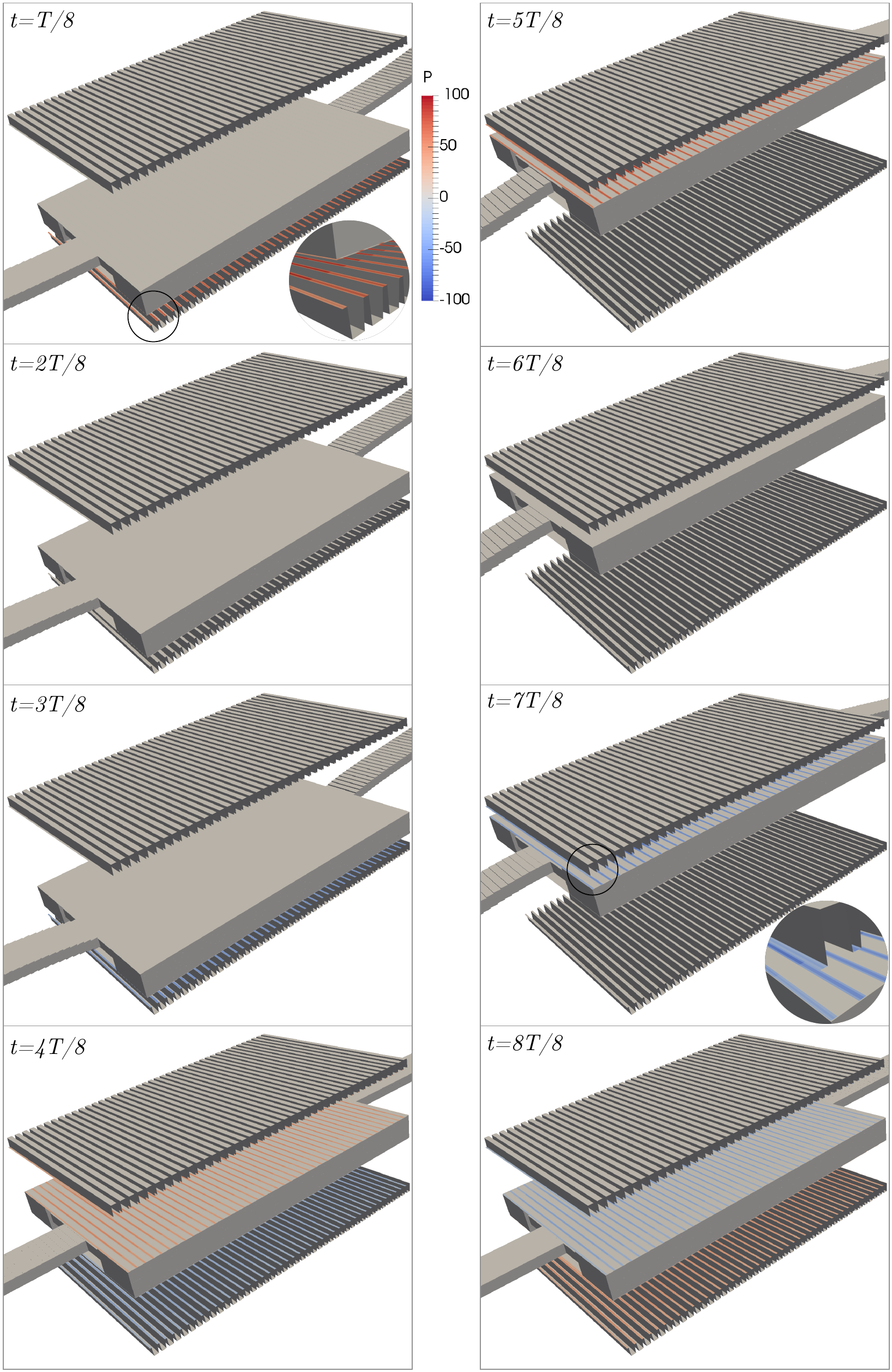}
\caption{Simulation results for case \textcolor{red}{WHAT?} over a complete cycle at the quasi-stationary state. The color denotes the departure from the average enclosure pressure in the squeeze film.}
\label{fig:seqP}
\end{figure}

In the simulation results presented next for the modified design, the
contact between the pyroelectric capacitor and the thermal reservoirs is modeled by imposing the corresponding limits on the displacement of the capacitor.
Figs. \ref{fig:Temp_LowStiff} and \ref{fig:Pos_LowStiff} show that, upon reducing the tether $x$-bending compliance and placing stops at the locations of the LTR and the HTR reservoirs, the operation cycle is nearly symmetric with a frequency of 75 $Hz$.  
Figures \ref{fig:seqT} and \ref{fig:seqP} show \textcolor{red}{WHAT?}

\section{Design Considerations}
\label{sec:designConsiderationsAndOptimization}
\extrafloats{100}

In this section, we discuss design considerations based on the time scales characterizing
heat transfer and structural dynamics. We also discuss the necessary conditions for sustained oscillation of the device, and answer the question of whether these sustained oscillations can take place at the mechanical
resonance frequency. Based on the discussion, we present in Appendix B a simple design procedure that yields preliminary dimensions which serve as starting values for the optimization algorithm wrapped around the model presented in Section \ref{sectionReducedOrderModel}.



\subsection{Heat Transfer Considerations}
Crucial to sustained oscillations of the pyroelectric capacitor is designing for the relation
between $\tau_c$, the time it takes the capacitor to store internal energy by heat transfer
from the HTR, and $\tau_s$, the time it takes the bimetal cantilevers to pull
the capacitor away for the reservoir. For sustained oscillations, the capacitor must store a sufficient amount of internal energy before it detaches from the HTR. After separation, the thermal energy stored in the capacitor continues to heat
the bimetal cantilevers which, in turn, exert a force on the capacitor that pushes it toward the LTR. The energy stored in the capacitor must be such that the heat gained
by the bimetal cantilevers is sufficient to push the capacitor all the way to the LTR. Similar arguments can be made for detachment from the LTR
and subsequent travel to the HTR.


The time scale $\tau_c$ is estimated by relating the capacitor's change in internal energy with the heat transferred by conduction through the thickness of the capacitor
\begin{equation}
\tau_{c}\ \sim\ \frac{t_{c}^2}{\alpha_c}
\label{eq:tau_Cap}
\end{equation}
where $t_{c}$ is thickness of the pyroelectric capacitor, and $\alpha_c \equiv \frac{k_c}{\rho_c c_c}$
is the thermal diffusivity of the pyroelectric material.  

In order for the capacitor to detach from the thermal reservoir, the force exerted by the
bimetal cantilevers on the capacitor should switch its direction so that it pulls the capacitor away
from the reservoir. How quickly this force reverse its direction depends on the change in
the bimetal cantilevers temperature. In estimating $\tau_{s}$, the time scale required for such a
change to take place, we assume that the bimetal cantilevers force needed to detach the capacitor
from the thermal reservoir develops
over the time it takes the heat from the capacitor, flowing through the tethers, to diffuse
along a fraction of the bimetal cantilever length, $L_b^*$. The resulting change in average
temperature of this portion (of length $L_b^*$) of the bimetal cantilever is  $\Delta T^*$. 
The time scale $\tau_{s}$ can then be estimated from the balance
\begin{equation}
\frac{T_c - T_b^*}{R_t + R_b^*} \tau_s = \rho_b c_b A_b L_b^* \Delta T^* \nonumber
\label{eq:tau_Sep}
\end{equation}
where $R_t = \frac{L_t}{4 k_t A_t}$ and $R_b^*=\frac{L_b^*}{2 k_b A_b}$ are respectively the thermal resistances
of the tethers and bimetal cantilevers of length $L_b^*$.
Note that introducing the tethers adds to the degrees of freedom so that we can
adjust $\tau_s$ by changing $L_t/(k_t A_t)$, provided that $(b/t)_t << 1$ and $(b/L)_t << 1$.
The tethers may be thought of as thermal resistors that can be designed to regulate
the flow of heat between
the pyroelectric capacitor and the bimetal cantilevers.






Hence, in order to maintain the device's oscillations the following condition should be
satisfied:


\begin{equation}
\frac{\tau_{c}}{\tau_{s}} << 1 \Rightarrow \frac{t_{c}^2}{L_b^{*2}} \frac{\alpha_b}{\alpha_c \left( 1 + \frac{R_t}{R_b^*}\right) } << 1
\label{eq:OscilationsCond}
\end{equation}
For the materials listed in Table \ref{table:materialProperties}, $\alpha_b/\alpha_c = 1.27$.
If $L_b^*$ is such that thermal resistance of the tethers is much less than that of the
bimetal cantilevers of characterisitc length $L_b^*$, then a necessary condition for the sustained cyclic operation of the device to be sustained is
\begin{equation}
t_c << L_b^* \mbox{ when } R_t << R_b^*
\label{eq:tetherHeatCond}
\end{equation}
Note that $L_b^*/L_b$ is primarily dependent on $\Delta T$ and on $H-t_c$, the travel distance
of the pyroelectric capacitor between the two reservoirs.

We point out when condition $R_t << R_b^*$ is satisfied, the time scale limiting the cycling frequency of the device is $\tau_{b}$; the time scale of thermal diffusion along the bimetal cantilever,
\begin{equation}
\tau_{b} \sim\ L_{b}^2  \bigg(\frac{\rho c}{k}\bigg)_{b}
\label{eq:tau_Bim_LB2}
\end{equation}


For the dimensions listed in Table  \ref{table:dimensions}, $\tau_c=1.1 \, \mu s$, $R_t/R_b=0.1$,  and $\tau_{b}=13.58 \, ms$.
Since $\tau_{b}$ is the time scale limiting the cycling operation of the device, the cycling frequency is of the order $1/\tau_b = 71 Hz$, which is close the frequency predicted by the simulations presented in Section \ref{sec:dynamicOperation}.


\subsection{Structural Mechanics Considerations}
As previously mentioned, structural mechanics considerations require the tethers to have a
small torsional stiffness and a bending stiffness in the $x-$direction that is much smaller than that of the bimetal cantilever.
This can be satisfied by choosing the tether width
$b_t$ to be as small as possible and selecting its length and thickness such that $b_t/L_t<0.15$,
 $t_t/b_t > 4$, and $t_t (b_t/L_t)^3 \ll b_b (t_b/L_b)^3$.
In order for the pyroelectric capacitor to travel the entire distance separating the
two reservoirs, the thermal expansion contribution to the force acting on the capacitor
must be larger than the restoring elastic component. Referring to Eq. (\ref{eq_ftipbimorpth}),
the length of the bimetal cantilever must be chosen to be sufficiently large. Note however, that choosing
$L_b$ to be too large results in reduction in the speed of the device. This is because $\tau_b
\propto L_b^2$ (see Eq. (\ref{eq:tau_Bim_LB2})). The pyroelectric capacitor thickness, $t_c$,
must be sufficiently large so that it behaves as a rigid body. Dimensions of the pyroelectric
capacitor should be carefully chosen as they also affect the thermal behavior (see Eq. (\ref{eq:tau_Cap})), the
pyroelectric effect and the resulting power output (Eqs. (\ref{EQwork}-\ref{EQpowerDensity})),
and the damping force (Eq. (\ref{eq:dampingForce})). 

\subsection{Operation at the Mechanical Resonance Frequency}
Referring to Eq. (\ref{EQpower}), the power output is proportional to the frequency,
volume of the pyroelectric capacitor, and the difference in temperature between the HTR and the LTR. Here we investigate the conditions of maximizing the
frequency for given pyroelectric capacitor volume and material properties of the
pyroelectric capacitor, tethers, and bimetal cantilever layers. We also answer the question whether
the device can operate at the mechanical resonance frequency. Increasing the operation
frequency can be accomplished by reducing the bimetal cantilever length
(Eq. (\ref{eq:tau_Bim_LB2})), which in turn requires decreasing the tether's length
and pyorelectric capacitor thickness, as required by conditions (Eq. \ref{eq:tetherHeatCond})
and (Eq. \ref{eq:OscilationsCond}). Due to the requirement of large bending compliance in
the $x-$direction, we expect the tethers width to be the smallest dimension in the device. Thus, the minimum dimension allowed by the fabrication process puts an upper bound on the maximum operation frequency of the device. In addition, reducing the pyroelectric capacitor thickness while keeping its volume constant requires increasing its planar area proportionally, which increases the gas damping force. In order for the device to operate at the mechanical resonance frequency, we require $f_{\mbox{res}} \tau_b \sim 1$ where the resonance frequency is
$f_{\mbox{res}} \simeq \frac{1}{2 \pi} \sqrt{\frac{2 \kappa_{b}}{m_c}}$ and the bending stiffness of
a single bimetal cantilever is $\kappa_b = \frac{3 (EI)_{eq}}{L_b^3}$. Note that, by design,
the contribution of the tethers to the stiffness is negligible. For the material properties
listed in Table \ref{table:materialProperties}, and choosing $b_1=b_2$ and
$t_2/t_1=\sqrt{E_1/E_2}$, the condition reduces to
\begin{equation}
9.3685 \, t_b \sqrt{V_b/V_c} \simeq 1,
\label{EqfresCond1}
\end{equation}
where $t_b=t_1+t_2$ is the total thickness of the bimetal cantilever in $\mu m$, and $V_b$ and $V_c$
are respecively the volumes of the bimetal cantilever and the capacitor.
It can then be shown that the resonance frequency, which in this case is the same as the
operating frequency, is related to the length ($L_b (\mu m)$) of the bimetal cantilever as
\begin{equation}
f_{\mbox{res}} = \frac{7.050 \times 10^7}{L_b^2}.
\label{EqfresCond2}
\end{equation}

In Appendix B, we present a design procedure to estimate the device dimensions so that it operates at the
mechanical resonance frequency for a given  $T_H-T_L$  and $H$. This procedure, along with the design considerations discussed above, were used to explore and explain the qualitative dependence on $T_H-T_L$ of the bimetal cantilevers length (Fig. \ref{fig:designLbfVSDT}),  operating frequency (which is equal to the mechanical resonance frequency) (Fig. \ref{fig:designLbfVSDT}), and the length of the pyroelectric capacitor (assumed to have a square
shape)  for different values of distance separating the two reservoirs (Fig. \ref{fig:designLcVSDT}) . These dependences for different values of the travel distance are also discussed and presented in Figs. \ref {fig:designLbVSDTVARDMIN}, \ref{fig:designfVSDTVARDMIN} and \ref{fig:designLcVSDTVARDMIN}.

We conclude by pointing that that the trends presented in Figs. \ref{fig:designLbfVSDT} - \ref{fig:designLcVSDTVARDMIN} are only qualitative since the  design procedure is not based on the dynamic solution of Eq. (\ref{eq:Newton}).
The procedure also does not include the tethers or the damping force, and as such, serves only to
gain physical insight and to provide initial values of the geometrical parameters to be used in the optimization algorithm that incorporates a numerical solver of 
the reduced order dynamic model presented in Section \ref{sectionReducedOrderModel}. The optimization algorithm is used to arrive at device dimensions that maximize the output work, as discussed in Section \ref{optimalDesigns}.

\section{Validation of the Model}\label{modelValidation}
Validation of the reduced order model was carried out in three phases. In the first phase, the heat transfer and the bimetal cantilever tether force models were validated by comparing with Ansys simulations. Refer to Appendix A for details.
In the second phase, validation of the coupled thermo-mechanical model was carried out by comparing with Ansys simulations the dynamic behavior of the device of dimensions listed in Table \ref{table:dimensions}. Other conditions are stated at the end of Section \ref{sec:dynamicOperation}.
The comparison presented in Figs. \ref{fig:Temp_LowStiff} and \ref{fig:Pos_LowStiff} for the change with time of the capacitor temperature and displacement over three operation cycles at the quasi-stationary state shows that the reduced order model accurately predicts the dynamic behavior of the device. 

\begin{figure}[!ht]
\centering
\centering
 \includegraphics[width = 4in]{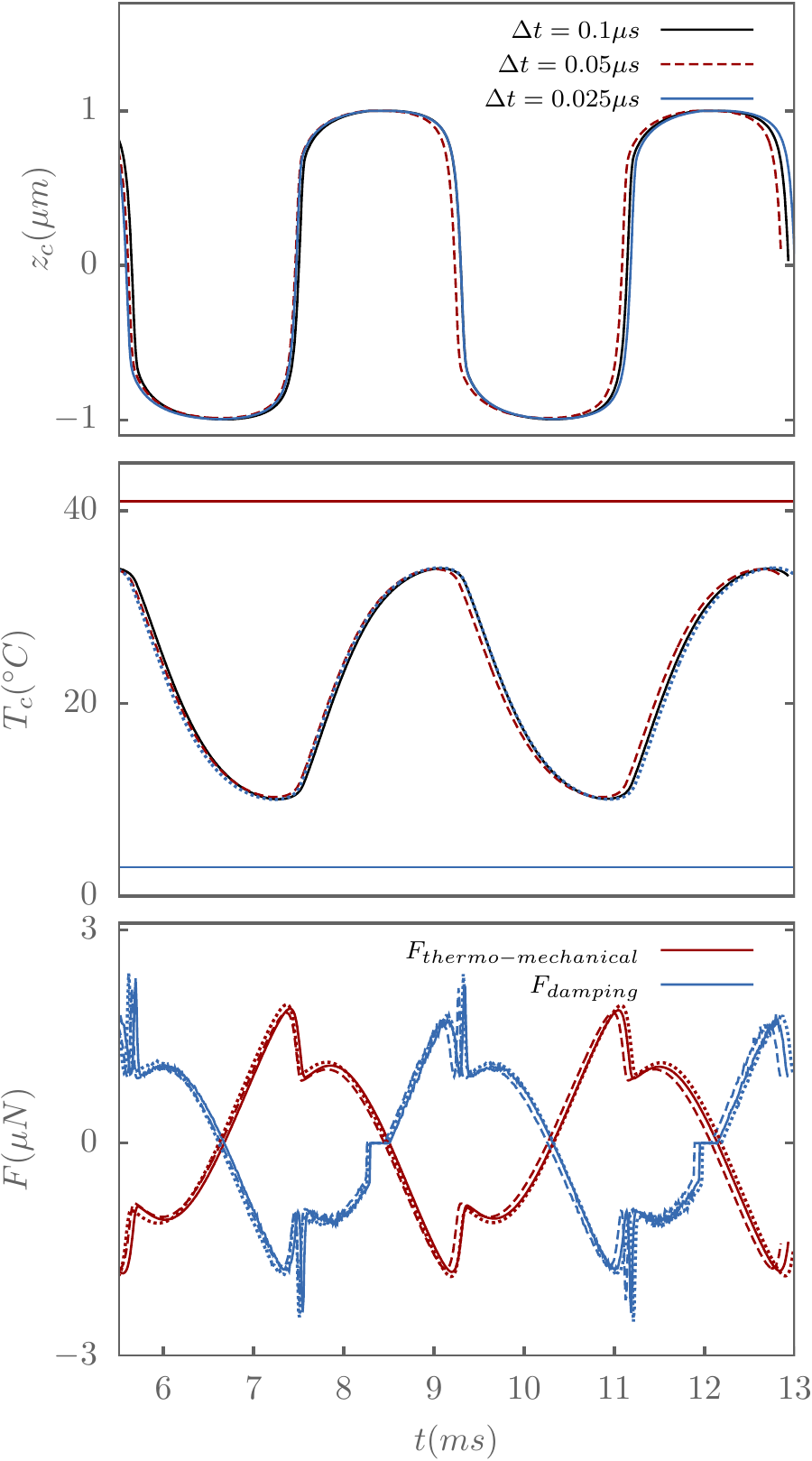}
\caption{Impact of the time step size on the convergence at quasi-stationary state.}
\label{fig:impactOfTimeStep}
\end{figure}

The third phase entails choosing the time step and spatial resolution used in the model as a tradeoff between accuracy and cost. 
Impact of the computational time step on the accuracy of the scheme is assessed for case 7 (Table \ref{table:OPTIMAL}) in terms of time evolution of the pyroelectric capacitor position, average temperature, and the damping and thermo-mechanical forces.  As can be seen from Fig. \ref{fig:impactOfTimeStep}, choosing $\Delta t = 0.1\,  \mu s$ yields a sufficiently accurate solution. Noting that for this case, the operating frequency is highest (291
$Hz$ from Fig. \ref{fig:fVSDT}), we chose $\Delta t = 0.1 \, \mu s$ for all the other cases. As for the mesh size, choosing $\Delta x = \Delta y = b_{tr}/10$ yielded a damping force that is nearly independent of the mesh size.

\section{Optimal Designs}\label{optimalDesigns}
The simple design procedure presented in Appendix B serves to provide the dynamic solver with initial
guesses of the various dimensions. The solver, which is based on the model presented in Section \ref{sectionReducedOrderModel},
simulates the thermo-mechanical dynamic operation, accounting for the heat transfer between the reservoirs, capacitor, and bimetal cantilevers, and for the thermo-mechanical, torsion and damping forces imparted on the capacitor respectively by the bimetal cantilevers, the tethers, and the surrounding fluid. Eq. (\ref{eq:Newton}) is numerically solved using a first order explicit time stepping scheme. Computing the damping force by integrating Eq. (\ref{eq:dampingForce}) requires solving Eq. (\ref{eqDampingPDE1}). Being the most expensive step, numerical solution of Eq. (\ref{eqDampingPDE1}) is carried out in parallel using OpenMP where the computational mesh is distributed among the CPU cores.

\begin{figure}[!ht]
\centering
\centering
 \includegraphics[width = 4in]{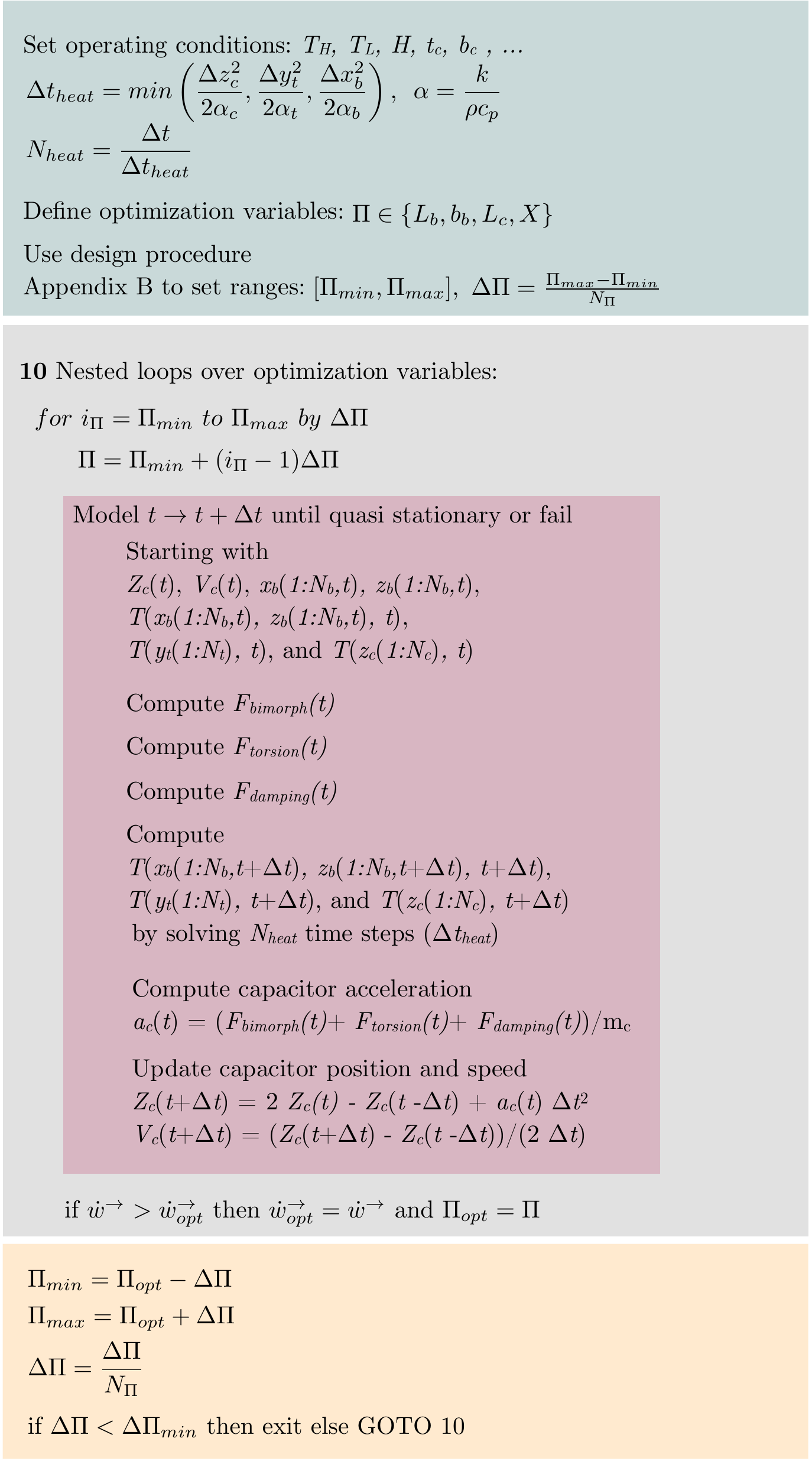}
\caption{Flowchart of the optimization algorithm and the embedded solver.}
\label{figAlgorithm}
\end{figure}

As shown in Fig. \ref{figAlgorithm}, the solver is embedded in an optimization algorithm that maximizes the work produced per unit volume (Eq. (\ref{EQpowerDensity})) while
meeting the various constraints imposed by the reduced order model. For example, all the optimal designs presented below satisfy the condition $\frac{E_b b_b (t_b/L_b)^3}{E_t t_t (b_t/L_t)^3} > 18$, which guarantees a sufficiently larger bending compliance (in the $x-$ direction) of the tethers.
The optimization algorithm
employs a basic incremental search along all the design coordinates. Once an improved design is identified, the incremental search
starting from the new improved design in repeated. The procedure continues until no further improvement is reached. We point
out here that gradient-based optimization approaches did not offer any advantages due to the fact that the solution space is
not only noncontinuous, but rather patchy. Nearly optimal designs were obtained using the optimization algorithm for different values of $T_H-T_L$. The various dimensions of these designs are listed in Table \ref{table:OPTIMAL}. 

\begin{table}[!ht]
\centering
\begin{tabular}{cccccccccccccccc}
\toprule
    case \# & $T_H$ & $T_L$ & $H$ & $d_{tr}$ & $X$ & $L_b$ & $b_b$ & $t_b$ & $L_c$ & $t_t$ & $b_t$ & $L_t$  \\
    \midrule
1 & 35 & 9 & 20 & 2.5 & 0.2893 & 454 & 13 & 4.25 & 1880  & 4.25 & 0.75 & 15 \\
2 & 36 & 8 & 20 & 2.5 & 0.3043 & 460 & 10 & 4.75 & 1880  & 5.75 & 0.75 & 15 \\
3 & 37 & 7 & 20 & 2.5 & 0.2850 & 460 & 9 & 5 & 1840  & 5 & 1 & 20 \\
4 & 38 & 6 & 20 & 2.5 & 0.3000 & 455 & 11 & 5 & 1880  & 5 & 1 & 20 \\
5 & 39 & 5 & 18 & 2.5 & 0.3029 & 453.75 & 13 & 5 & 1880  & 5 & 1 & 20 \\
6 & 40 & 4 & 16 & 2.5 & 0.2981 & 460 & 15 & 5 & 1880  & 5 & 1 & 20 \\
7 & 41 & 3 & 14 & 2.25 & 0.2156 & 435 & 11 & 5 & 1900  & 5 & 1 & 20
    \\
    \bottomrule
\end{tabular}
\caption{Dimensions of optimized designs for different $T_H-T_L$.
For all the designs, the width of the capacitor, minimum gap size, trench width were fixed: $b_c = 50 \, \mu m$, $g_{min} = 0.5 \, \mu m$, $b_{tr}=25 \, \mu m$, $X = 1 - b_{tr}/a$, $t_c=H-2\, \mu m$. Note that the smallest dimension is the tether width $b_t$, with $0.75 \, \mu m \leq b_{t} \leq 1 \, \mu m$. All dimensions are in $\mu m$ and temperatures in $\degree C$.
}
\label{table:OPTIMAL}
\end{table}

\begin{figure}[!ht]
\centering
\centering
 \includegraphics[width = 4.5in]{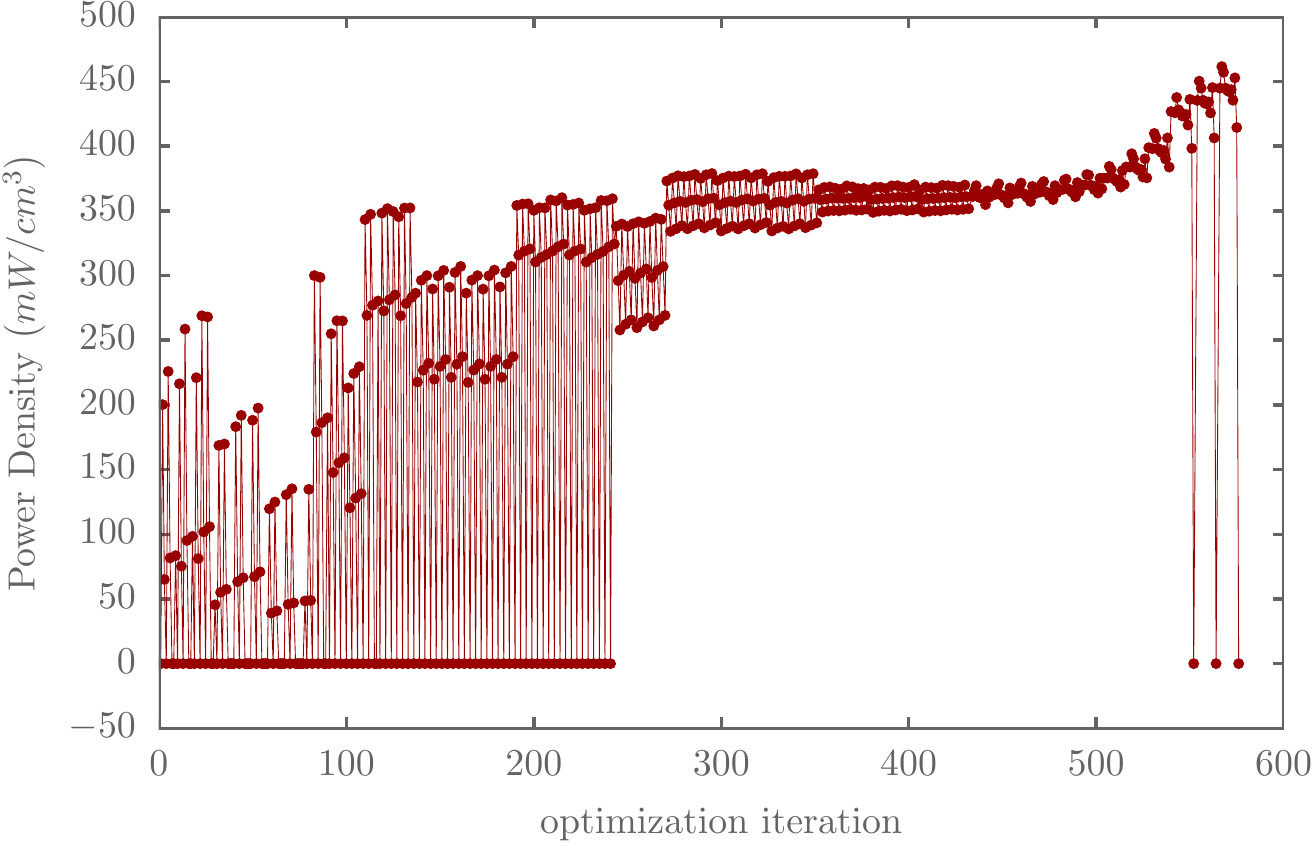}
\caption{An example showing the power density versus the optimization iteration for the case \textcolor{red}{HERE}.}
\label{fig:optim}
\end{figure}

Figure \ref{fig:optim} shows  \textcolor{red}{HERE}

\begin{figure}[!ht]
\centering
\centering
 \includegraphics[width = 4in]{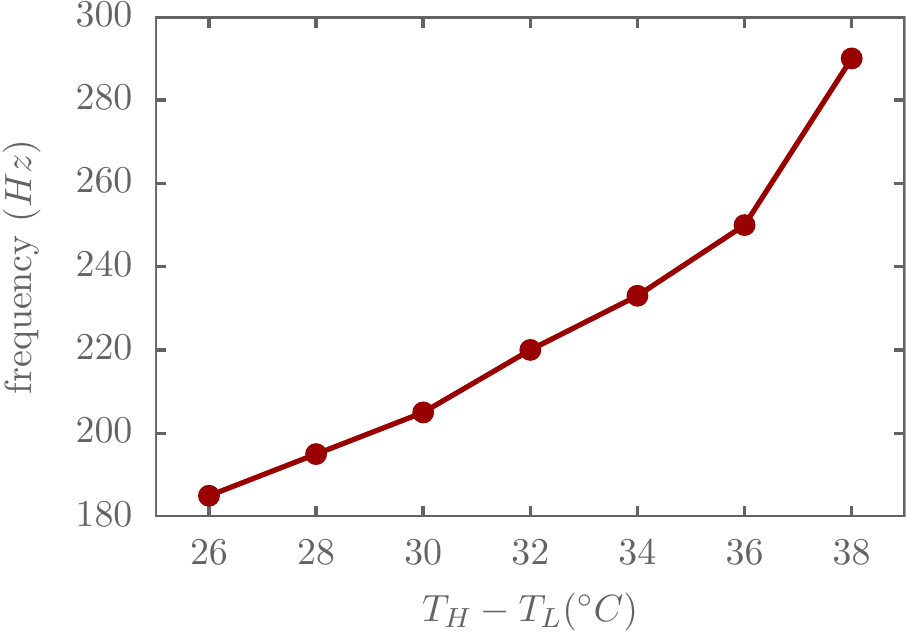}
\caption{Dependence of the operating frequency on the difference in temperature between the HTR and the LTR for the cases listed in Table \ref{table:OPTIMAL}.}
\label{fig:fVSDT}
\end{figure}

\begin{figure}[!ht]
\centering
\centering
 \includegraphics[width = 4in]{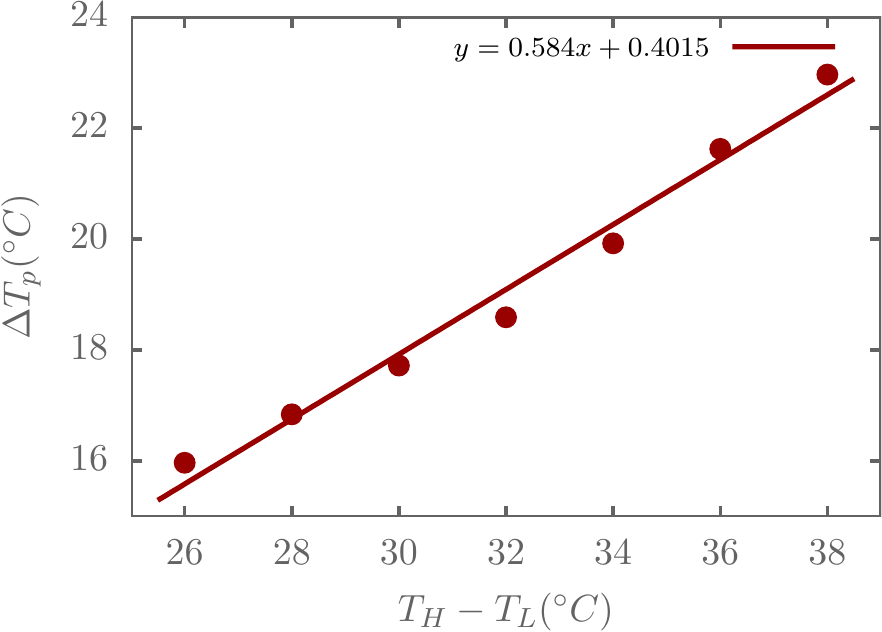}
\caption{Dependence of the peak to peak change in the average capacitor temperature over a cycle on the difference in temperature between the HTR and the LTR for the cases listed in Table \ref{table:OPTIMAL}.}
\label{fig:DTpVSDT}
\end{figure}

\begin{figure}[!ht]
\centering
\centering
 \includegraphics[width = 4in]{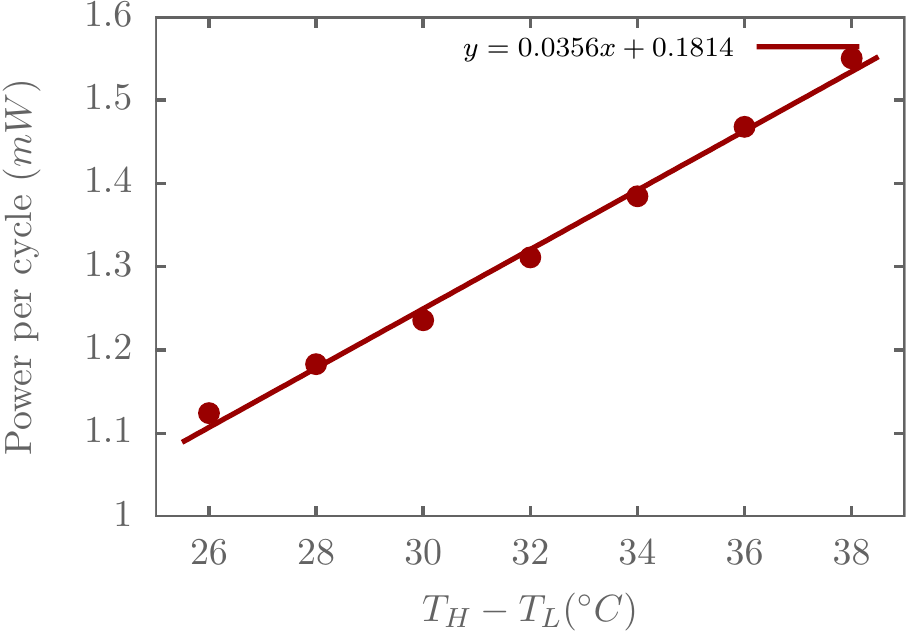}
\caption{Dependence of power generated per cycle on the difference in temperature between the HTR and the LTR for the cases listed in Table \ref{table:OPTIMAL}.}
\label{fig:PowerOfVSDT}
\end{figure}

\begin{figure}[!ht]
\centering
\centering
 \includegraphics[width = 4in]{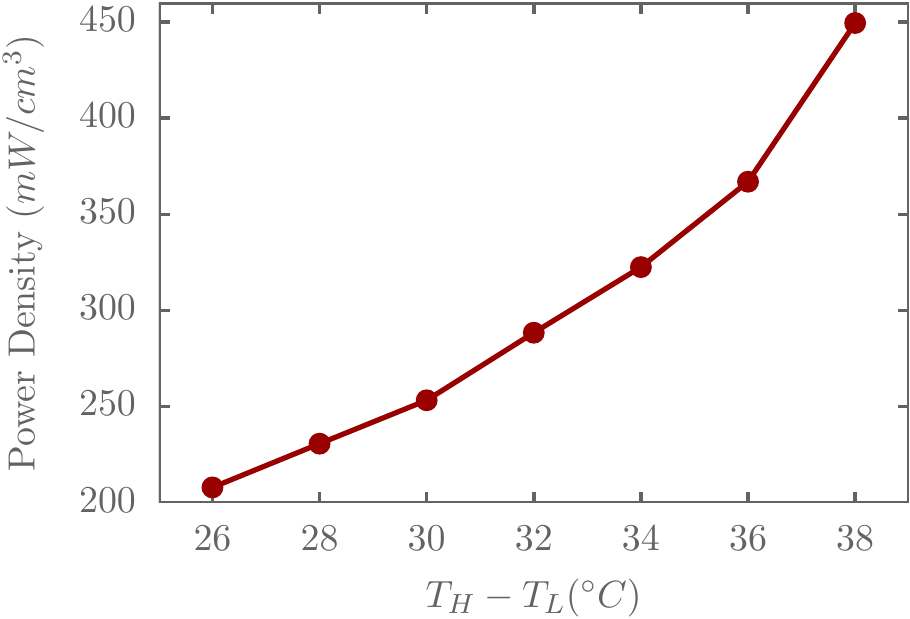}
\caption{Dependence of power generated per unit volume on the difference in temperature between the HTR and the LTR for the cases listed in Table \ref{table:OPTIMAL}.}
\label{fig:powerVSDT}
\end{figure}

As shown in Fig. \ref{fig:fVSDT},  the operation frequency increases with temperature difference between the two reservoirs,  in accordance with the discussion in Section \ref{sec:designConsiderationsAndOptimization}. The operation frequency ranges from 183 $Hz$ for $T_H-T_L=26 \degree C$ to 291 $Hz$ for $T_H-T_L=38 \degree C$.
Since the pyroelectric capacitor does not get in contact with either reservoir but rather reaches within a distance of  $g_{min}$, we expect that, $\Delta T_p$, the peak to peak change in the average capacitor temperature over a cycle to be less than $T_H-T_L$. The plot presented in Fig. \ref{fig:DTpVSDT} show that $\Delta T_p$ increases linearly with $T_H-T_L$, with the ratio $\Delta T_p/(T_H-T_L)$ being nearly a constant of $\sim$ 0.6.
The power produced per cycle, $\dot{\omega}^{\rightarrow}/f$, is plotted against $T_H-T_L$ in Fig. \ref{fig:PowerOfVSDT}. It follow a linear trend similar to $\Delta T_p$, which follows from Eq. (\ref{EQpowerDensity}).
The power density per unit volume, expressed in Eq. (\ref{EQpowerDensity}), increases quadratically in $T_H-T_L$, as seen in the plot of Fig. \ref{fig:powerVSDT}. As $T_H-T_L$ increases from 26 to $38 \degree C$, the power density increases from $208$ to $450 \, mW/cm^3$.  

We finally note that, for the considered range of $T_H-T_L$, the power generated per unit surface area ranges from 0.4 to 0.65 $mW/cm^2$, as presented in Fig.  \ref{fig:powerPerAreaVSDT} of Appendix C. These values are compared with power density values of energy harvested from other environmental sources in Table \ref{table:energyHarvesting}.

\begin{table}[!htbp]
\centering
\resizebox{\columnwidth}{!}{
\begin{tabular}{|c|c|c|c|c|}
\hline
\bf{Energy source} & \bf{Method} & \bf{Characteristics} & \bf{Harvested Power} & \bf{Efficiency}\\
 & & &  \bf{Density} ($mW/cm^2$) &  \\
\hline
Light & Photovoltaic & Outdoor & 10 & 5-30 \%\\
         &                      & Indoor  & 0.01 & \\
\hline
Ambient air flow & Microturbines  & & 1  & \\
\hline
Mechanical vibration & Piezoelectricity & machine ($kHz$) & 0.1-0.8 & 1-10 \% \\
 &  &  human ($Hz$) & $0.004$ &\\
 \hline
Mechanical vibration & Electromagnetic & machine ($kHz$) & 2 & \\
 &  &  human ($Hz$) & $0.05$ &\\
 \hline
Thermal & Thermoelectric & machine  & 1-10 & 0.15 - 10 \%\\
 &  &  human  & $0.025-0.06$ & \\
\hline
{\it Thermal} & \it{Pyroelectric} & $\Delta T$ = 26-38  \degree C&  0.4 - 0.65 &  5.4-6.74\%\\
              & (This work) & & & \\
 \hline
RF & Electromagnetic & background & (0.001-0.1) $\times 10^{-3}$  & \\
     &                            & directed &  1       & 33 \% \\
\hline
\end{tabular}
}
\caption{Power density of energy harvested from various environmental sources\cite{Rasheduzzaman2016ASO,VULLERS2009684,1401839}. The energy harvested using the pyroelectric generator presented in this paper is also listed.}
\label{table:energyHarvesting}
\end{table}%

\begin{figure}[!ht]
\centering
\centering
 \includegraphics[width = 4in]{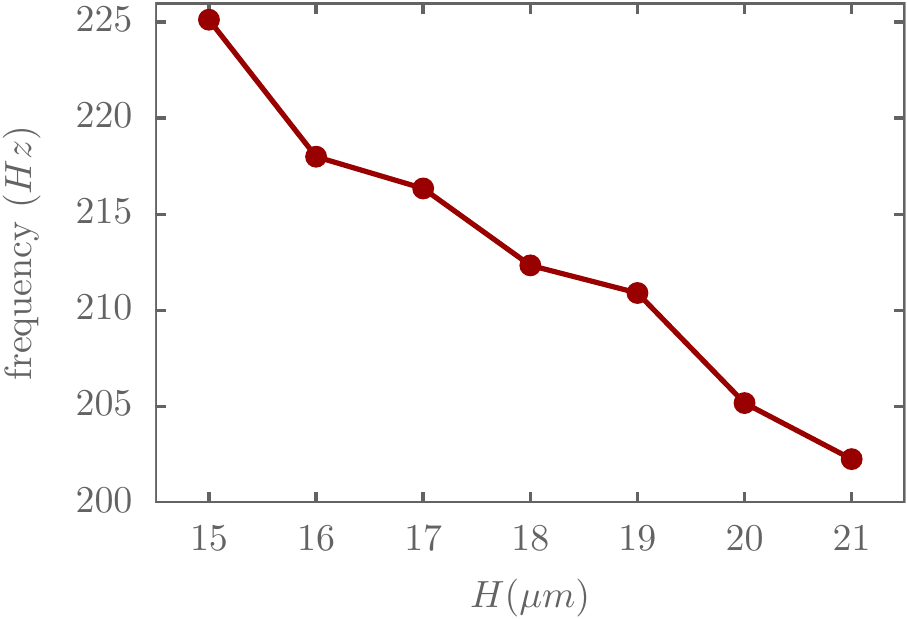}
\caption{Variation of the operating frequency with the distance separating the two thermal reservoirs for the cases listed in Table \ref{table:OPTIMAL2}.}
\label{fig:fVSH0}
\end{figure}

\begin{figure}[!ht]
\centering
\centering
 \includegraphics[width = 4in]{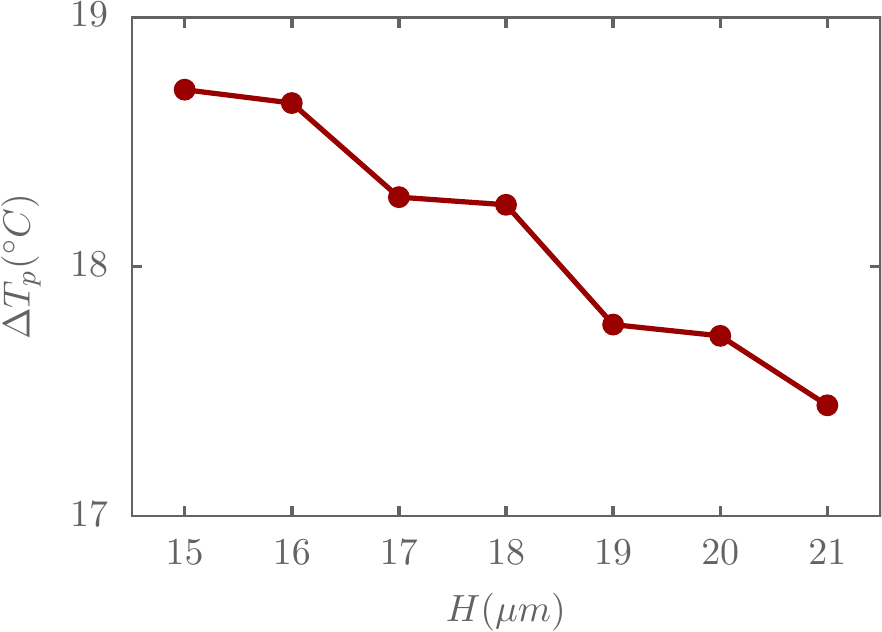}
\caption{Variation of the peak to peak change in the average capacitor temperature over a cycle  with the distance separating the two thermal reservoirs for the cases listed in Table \ref{table:OPTIMAL2}.}
\label{fig:DTpVSH0}
\end{figure}

\begin{figure}[!ht]
\centering
\centering
 \includegraphics[width = 4in]{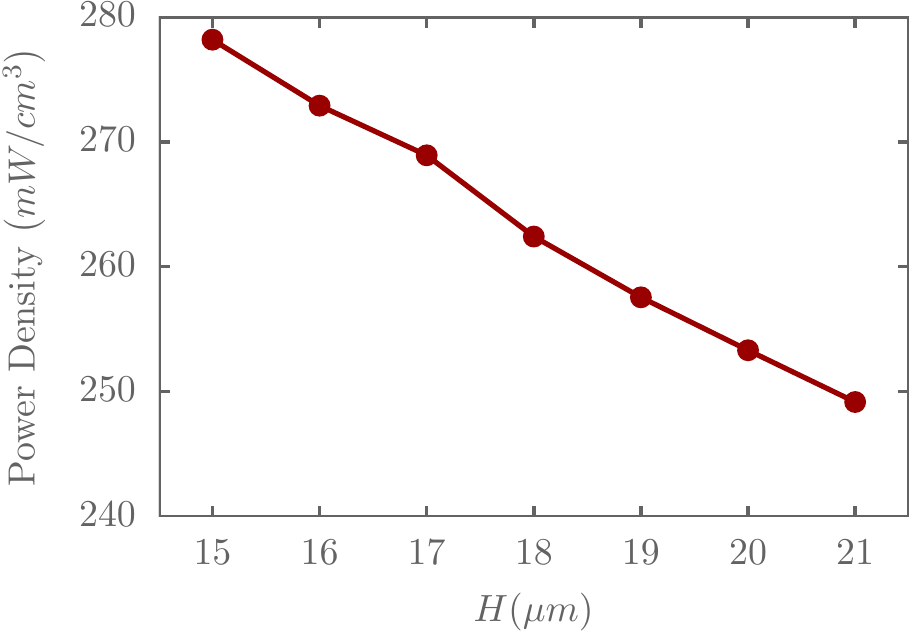}
\caption{Variation of power generated per unit volume with the distance separating the two thermal reservoirs for the cases listed in Table \ref{table:OPTIMAL2}}
\label{fig:powerDensityH0}
\end{figure}

Next, we investigate impact of $H$, the separation between the HTR and the LTR. The temperatures of the HTR and the LTR are kept at $ 37$ and $7$ $\degree C$ respectively. The thickness of the pyroelectric capacitor is selected to be $t_c = H - 2 \, \mu m$. Other dimensions are listed in Table \ref {table:OPTIMAL2}.
\begin{table}[!ht]
\centering
\begin{tabular}{cccccccccccccccc}
\toprule
    case \# & $H$ & $d_{tr}$ & $X$ & $L_b$ & $b_b$ & $t_b$ & $L_c$ & $t_c$ & $t_t$ & $b_t$ & $L_t$  \\
    \midrule
1 & 15 & 3.5 & 0.305 & 480 & 10 & 5 & 1880  &  13 & 5 & 1 & 20 \\
2 & 16 & 3.5 & 0.305 & 475 & 10 & 5 & 1880  &  14 & 5 & 1 & 20 \\
3 & 17 & 3 & 0.305 & 470 & 10 & 5 & 1880  &  15 & 5 & 1 & 20 \\
4 & 18 & 2.75 & 0.300 & 465 & 9 & 5 & 1840  &  16 & 5 & 1 & 20 \\
5 & 19 & 2.75 & 0.300 & 465 & 9 & 5 & 1800  &  17 & 5 & 1 & 20 \\
6 & 20 & 2.5 & 0.285 & 460 & 9 & 5 & 1840  &  18 & 5 & 1 & 20 \\
7 & 21 & 2.5 & 0.305 & 460 & 9 & 5 & 1880  &  19 & 5 & 1 & 20 \\
    \bottomrule
\end{tabular}
\caption{Dimensions of optimized designs for different $H$.
For all designs, $T_H = 37 \degree C$ , $T_L = 7 \degree C$, $b_c = 50 \, \mu m$, $g_{min} = 0.5 \, \mu m$, $X = 1 - b_{tr}/a$,
$b_{tr}=25 \, \mu m$, and $t_c = H - 2 \, \mu m$. All dimensions are in $\mu m$.
}
\label{table:OPTIMAL2}
\end{table}
Fig. \ref{fig:fVSH0} shows that as $H$ increases, the operating frequency decreases. This decrease is, however, small because the gap separating the capacitor from either reservoir is kept constant by adjusting the capacitor thickness according to $t_c = H - 2 \, \mu m$. As such, the distance traveled per cycle is the same for all cases and the slight change in frequency is attributed to the combined effect of a thicker capacitor and shorter bimetallic beams. As $H$ increases from 15 to 21 $\mu m$, the peak to peak change in the average capacitor temperature over a cycle decrease by $\sim$1.2 $\degree C$, according to Fig.  \ref{fig:DTpVSH0}. Since both $f$ and  $\Delta T_p$ decrease as $H$ is increased, the power generated per unit volume decreases by $\sim 10 \%$ , as shown in Fig. \ref{fig:powerDensityH0}. 

\section{Conclusion}
In this study, a new design of a MEMS pyroelectric energy harvester is proposed. During cyclic operation of the device, the pyroelectric capacitor plate remains parallel to the thermal reservoirs, which not only increases the rate of heat transfer across the gap, but also ensures nearly uniform temperature within the capacitor. This operation is enabled by the use of tethers of low torsional stiffness to connect the bimetal cantilevers to the plate. Investigation of various times scales key to thermo-mechanical operation of the device uncovered the conditions for self-sustained oscillations, and the conditions for these oscillations to take place at the mechanical resonance frequency (in the absence of gas damping).  The physically-based reduced order model was employed to identify the length of the bimetal cantilever beams as a key design parameter. When embedded within an optimization algorithm, the model was used to arrive at device dimensions for nearly optimal power generation while operating in the self-sustained oscillations/contact-free operation regime. As the temperature difference between the two reservoirs increases from 26 to 38 $\degree C$, the power density harvested increased from  0.4 to 0.65 $mW/cm^2$.

The proposed design along with the design methodology presented fill a gap in the under-explored area of MEMS pyroelectric energy harvesters, which brings us one step closer to realization of these devices.

\section*{Acknowledgments}
This work was supported by the Munib and Angela Masri Institute of Energy and Natural Resources, Award Number 103352. 
The authors would like to thank Prof. Daniel Tartakovsky for his insightful comments.

\newpage

\bibliographystyle{iopart-num}
\bibliography{biblio}



\end{document}